\documentclass[noshowpacs,prl,aps,superscriptaddress,floatfix,twocolumn,nofootinbib]{revtex4-2} 
\usepackage[utf8]{inputenc}
\usepackage{lineno,mathtools,url}
\usepackage{amsmath}
\usepackage{soul}
\usepackage{amssymb,mathrsfs}
\usepackage{physics}
\usepackage{graphicx}
\usepackage{hyperref}

\hypersetup{
     colorlinks=true,
     linkcolor=blue,
     filecolor=blue,
     citecolor = orange,      
   }
\usepackage[dvipsnames]{xcolor}
\usepackage{dsfont}
\usepackage{mathbbol} 
\usepackage[normalem]{ulem}
\allowdisplaybreaks
\begin{document}

\title{
Oscillating Solitons and AC Josephson Effect in Ferromagnetic Bose-Bose Mixtures
}

\author{S. Bresolin}
\affiliation{Pitaevskii BEC Center, CNR-INO and Dipartimento di Fisica, Universit\`a di Trento, Via Sommarive 14, 38123 Povo, Trento, Italy}
\author{A. Roy}
\affiliation{Pitaevskii BEC Center, CNR-INO and Dipartimento di Fisica, Universit\`a di Trento, Via Sommarive 14, 38123 Povo, Trento, Italy}
\affiliation{School of Physical
Sciences, Indian Institute of Technology Mandi, Mandi-175075 (H.P.), India}
\author{G. Ferrari}
\affiliation{Pitaevskii BEC Center, CNR-INO and Dipartimento di Fisica, Universit\`a di Trento, Via Sommarive 14, 38123 Povo, Trento, Italy}
\author{A. Recati}
\email[]{Corresponding Author: alessio.recati@ino.cnr.it}
\affiliation{Pitaevskii BEC Center, CNR-INO and Dipartimento di Fisica, Universit\`a di Trento, Via Sommarive 14, 38123 Povo, Trento, Italy}
\author{N. Pavloff}
\affiliation{Universit\'e Paris-Saclay, CNRS, LPTMS, 91405, Orsay, France}
\affiliation{Institut Universitaire de France (IUF)}
\begin{abstract}
Close to the demixing transition, the degree of freedom associated with relative density fluctuations of a two-component Bose-Einstein condensate is described by a nondissipative Landau-Lifshitz equation. 
In the quasi-one-dimensional weakly immiscible case, this mapping surprisingly predicts that a dark-bright soliton should oscillate when subject to a constant force favoring separation of the two components. We propose a realistic experimental implementation of this phenomenon which we interpret as a spin-Josephson effect in the presence of a movable barrier.
\end{abstract}

\maketitle

Periodic motion under the effect of a uniform force field is a counterintuitive phenomenon occurring in some peculiar dissipation-less quantum-mechanical systems.
The most well-known example is represented by Bloch oscillations of a particle in a periodic potential 
\cite{bloch1929}, which are due to the wave nature of particles and the consequent energy band structure.
Another remarkable example due to quantum coherence is the AC Josephson effect, where 
a fixed voltage induces an oscillating current across a superconducting junction.
Such an effect 
also exists in other systems which break a continuous symmetry \cite{beekman2020}. In particular it occurs in superfluid $^3$He \cite{Avenel1988,Pereverzev1997} and $^4$He \cite{Sukhatme2001} and
in systems exhibiting Bose-Einstein condensation (BEC), such as ultracold gases 
\cite{Cataliotti2001,Albiez2005,Levy2007}, magnons \cite{kouki2014} and exciton-polaritons \cite{lago2010}. 
Two weakly coupled ferromagnets or antiferromagnets can also show the AC Josephson effect for the spin current in a mechanism referred to as the spin-Josephson effect, see, e.g., \cite{Markelov1987,Borovik1988,Nogueira2004,Nogueira2004,chen_macdonald_2017}. 

A different instance of oscillatory motion under a dc drive concerns certain solitons in Galilean-invariant systems. To our knowledge, such behavior was first discussed in \cite{Kosevich1998,Kosevich2001}, in the context of solitonic solutions of the dissipationless Landau-Lifshitz  equation (LLE), which describes the nonlinear dynamics of the local spin in a ferromagnet \cite{Landau_stat2}.
Very recently, similar dynamics have been found for two solitonic solutions in spinor condensates in ultracold gases:
a magnetic soliton in a two-component BEC with very specific interaction strengths \cite{Zhao2020} and a ferro-dark soliton in the ferromagnetic phase of a spin-1 BEC \cite{xiaoquan22}.
Furthermore, it has been shown in \cite{Schecter2012,Meinert2017} that a single impurity in a zero-temperature one-dimensional Bose gas also exhibits a peculiar damped oscillating dynamics under a constant force.

In the LLE, Kosevich and collaborators attributed the strange dynamics to the periodic dispersion relation ? as for Bloch oscillations ? and to the stability under an external uniform magnetic field of the easy-axis magnetic solitons. 
In the case of spinor condensates, the reason for the numerically observed dynamics was related to the oscillation between two solitonic solutions with positive and negative mass \cite{Zhao2020, xiaoquan22}.
Finally, for the impurity in the one-dimensional Bose gas the explanation of the periodic motion was based on the impurity cutting the gas and behaving as the barrier of a mobile Josephson junction \cite{Schecter_2016}, and on Bragg reflection induced by the strong bath correlation and the characteristic 1D spectrum \cite{Meinert2017}. 

\begin{figure}[h]
    \centering
    \includegraphics[width=0.95\linewidth]{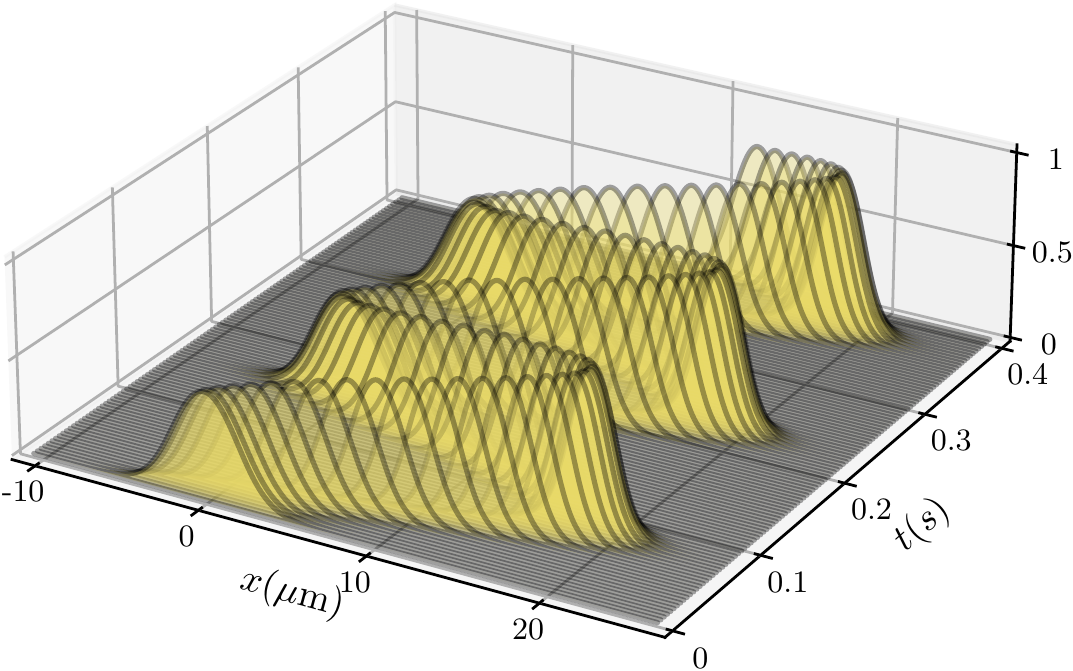}
    \caption{Numerical results for the evolution under a constant force of a dark-bright soliton in an immiscible mixture condensate of two hyperfine states of Na ($\ket{F=1,m_F=-1}$ and $\ket{F=2,m_F=-2}$). We represent the density of the minority component (in arbitrary units) as a function of position and time.}
    \label{fig:fig1}
\end{figure}

In the present Letter we first exploit a mapping between the Bose mixture and a ferromagnetic system to give a unified interpretation of this phenomenon alternative to previous ones \cite{Kosevich2001,Zhao2020}: we argue that, as for the single impurity model of Ref. \cite{Schecter_2016}, the oscillations of the soliton in the presence of a constant force
? such as represented in Fig. \ref{fig:fig1} ? are due to an unconventional Josephson effect. 
This interpretation suggests that the phenomenon is not restricted to the exact solitonic solution or to regimes where the mapping between a two-component BEC and a ferromagnet is valid. In the second part of the Letter we show that, indeed, the oscillating dynamics under a constant force is a more general feature of small spin domains. The breaking of integrability is reflected in nonperfectly sinusoidal oscillations of the spin domain. However, the majority component current preserves its sinusoidal character, as expected for a Josephson current. 
Such a robustness implies that it should be possible to (i) directly observe oscillating dynamics in a Galilean-invariant system using present technology in cold gas platforms (thus contributing to settle the controversy concerning the possible observation of this phenomenon \cite{Gamayun2014,Schecter2015_com,Gamayun2015_com})
and (ii) realize the analog of the voltage-current characteristic of a superconducting Josephson junction (SJJ).
So far, indeed, the Josephson effect in BEC has been related to the coherent relative density oscillations between two weakly linked condensates, either in double well traps or in two hyperfine levels \cite{Cataliotti2001,Albiez2005,Zibold2010}. Such a dynamics is described by the so-called Bose-Josephson junction equations, i.e., nonrigid pendulum equations \cite{Smerzi1997,Raghavan1999}, which interestingly show some new phenomena not observable with SJJs.
The magnetic soliton, or more generally the magnetic domain under the external potential, instead realizes a perfect analogue of the AC SJJ (see also \cite{Giovanazzi2000}).

Our platform is a two-component Bose gas at zero temperature. The mixture is physically realised by properly populating two hyperfine states of the atomic species of mass $m$ forming the gas.
The system is well described as a BEC with a spinor order parameter $\Psi(\vec{r},t)=(\psi_1\;\psi_2)^T$ obeying a Gross-Pitaevskii equation:
\begin{equation}\label{GP}
i\,\hbar\, \partial_t \Psi
=\left(-\frac{\hbar^2}{2 m}\Delta
+V_{\rm ext} + U_{\rm mf}\right) \Psi\; ,
\end{equation}
with
\begin{equation}
V_{\rm ext}=\begin{pmatrix}
V_1 & 0\\
0 & V_2
\end{pmatrix},\;
U_{\rm mf}=
\begin{pmatrix}
g_{11}\, |\psi_1|^2    &g_{12} \psi_2^*\psi_1 \\
g_{12} \psi_1^*\psi_2 &g_{22}\, |\psi_2|^2
\end{pmatrix},
\end{equation}
where $V_i(\vec{r}\,)$ is an external potential acting on component $i$ ($i=1$ or 2) and $g_{ij}$ are the positive intra- ($i=j$) and inter- ($i\neq j$) species interaction strengths. 
In a homogeneous configuration, i.e., $V_{\rm ext}\equiv 0$, the system exhibits a first-order phase transition from a miscible to an immiscible state depending on 
the relative value of the $g_{ij}$'s. Within the Gross-Pitaevskii description the system is miscible as long as $g_{12}<\sqrt{g_{11}g_{22}}$ (see, e.g., Ref. \cite{Pitaevskii2016}). 

Our goal is to describe quasi-one-dimensional configurations, so we consider the system to be  confined in an elongated geometry in order for the dynamics of the gas to occur only in the $x$ direction. 
The Gross-Pitaevskii equation can be conveniently recast in the form of spin superfluid hydrodynamics (see, e.g., \cite{Nikuni2003}) for the total density $n=\Psi^T\cdot\Psi$ and the spin density $\vec{s}=\Psi^T\vec{\sigma}\,\Psi$, where $\vec{\sigma}$ is the vector of Pauli matrices.   
The spin superfluid nature of BEC mixtures, collective spin modes, the role of the SU(2) symmetry breaking (due to the nonequality of the $g_{ij}$'s as well as to the presence of an external transverse magnetic field) have recently received important experimental verifications \cite{Kim2017,Fava2018,Lepoutre2018,Farolfi2020,Farolfi2020,Kim2021,Cominotti2022}.

As already discussed in 
Refs. \cite{Son2002,Qu2016,Congy2016,Ivanov2017}, the density and spin degrees of freedom essentially decouple close to the defocusing Manakov regime \cite{Manakov1974}, in the limit
\begin{equation}\label{decouple}
    |g_{11}-g_{22}| \;\; \mbox{and}\;\; |g_s| \ll g\; ,
\end{equation}
where $g=(g_{11}+g_{22})/2$ and $g_s=g_{12}-g$. The parameter $g_s$ ($\ne 0$) can be seen as an effective spin interaction;
it provides the natural units of length $\xi_s \equiv \hbar/\sqrt{2m n_0\abs{g_s}}$ and time $\tau_s \equiv \hbar/(n_0\abs{g_s}) $ for the spin dynamics in a system with homogeneous density $n_0$. In the regime \eqref{decouple}, using the rescaled variables $x/\xi_s \rightarrow x$ and $t/\tau_s \rightarrow t$, the equation of motion for the  magnetization $\vec{M}=\vec{s}/n$ can be written in the form of a one-dimensional dissipationless LLE \cite{Ivanov2017}:
\begin{equation}
 \label{LLE}
 \partial_t \vec{M} = (\vec{H}_{\textrm{eff}} + \vec{H}_{\textrm{ext}}) \wedge \vec{M},
\end{equation}
where $\vec{H}_{\rm ext} = \omega_D\vec{e}_z$ is an external field, $\vec{H}_{\textrm{eff}} \equiv \partial_x^2 \vec{M} + \epsilon M_z \vec{e}_z$, and we introduced the dimensionless quantities
\begin{equation}
\omega_D \equiv \frac{V_1-V_2}{\abs{g_s} n_0}, \;\; \epsilon \equiv \frac{g_s}{\abs{g_s}},
\end{equation}
where the adimensional differential potential $\omega_D$ can, in general, depend on position. For $\epsilon = -1$ ($\epsilon=+1$), corresponding to a slightly miscible (immiscible) mixture, Eq. \eqref{LLE} describes the evolution of the magnetization vector in an easy-plane (easy-axis) ferromagnet.
The relevance of the LLE for describing the dynamics of elongated BEC mixtures has recently been experimentally addressed in \cite{Farolfi2021}.

We now turn our attention to magnetic solitons.   
When $H_{\textrm{ext}}$ is constant, Eq. \eqref{LLE} is exactly integrable and its solitonic solutions are known \cite{Kosevich1990}. The equivalence with Eq. \eqref{GP} yields analytic expressions for the spin-solitonic solutions of the coupled Gross-Pitaevskii equations in the near-transition BEC mixture. In the following we restrict our attention to the immiscible situation (easy-axis LLE) such as considered e.g., in Ref. \cite{Chai2022} and defer a discussion of the miscible case to the Supplemental Material \cite{supplemental}. 
It is convenient to write $\vec{M} \equiv -(\sin\theta\cos\varphi, \; \sin\theta\sin\varphi, \; \cos\theta)^{T}$, corresponding to the parametrization \cite{Matthews1999}
\begin{equation}\label{madelung}
\Psi=
\begin{pmatrix}
\sqrt{n_1}\,e^{i\phi_1} \\[1mm]
\sqrt{n_2}\,e^{i\phi_2}  \\
\end{pmatrix}
=
\sqrt{n}\, e^{i\Phi/2}
\begin{pmatrix}
\cos\frac{\theta}2\,e^{-i\varphi/2} \\[1mm]
\sin\frac{\theta}2\,e^{i\varphi/2}  \\
\end{pmatrix}\; .
\end{equation}
The solitons are characterized by two parameters: the conserved quantity associated with the total $z$ magnetization $N= \int (1-\cos\theta) \dd{x}$ (in the Bose-mixture language $N=2N_2/n_0\xi_s$, where $N_2$ is the number of atoms of the minority component) and the total (adimensional) momentum $P = \int \partial_x\varphi (1-\cos\theta) \, \dd x$. In terms of these quantities, the soliton energy reads
\begin{equation}
\label{imm_energy}
E_{\rm sol}
= 4\tanh(N/4) + 8\, \frac{\sin^2(P/4)}{\sinh(N/2)}.
\end{equation}
\begin{figure*}[t!]
    \centering
    \includegraphics[width=\textwidth]{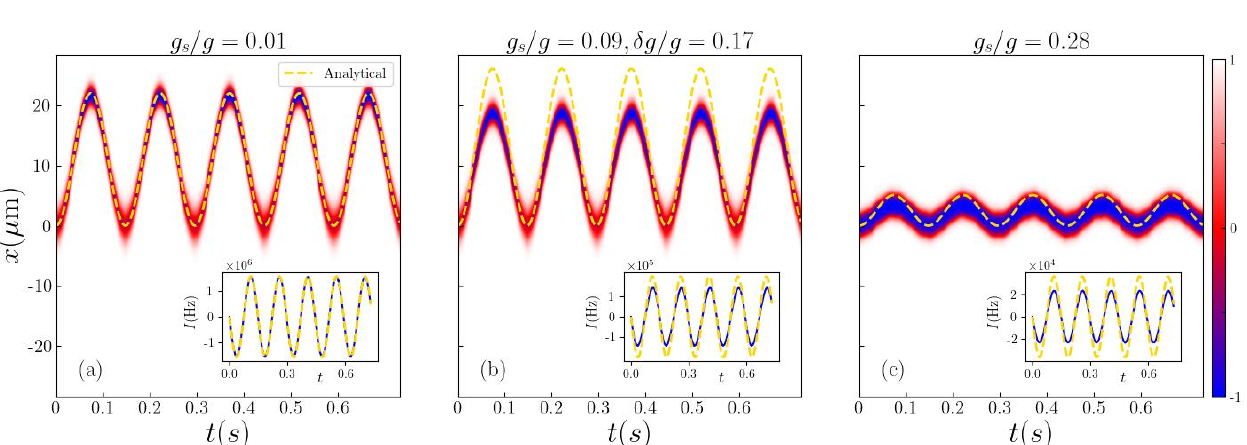}
    \caption{Numerical evolution of a spin domain under a linear differential potential for different values of the interaction strengths; (a) $g_s/g=0.01$, (b) $g_s/g=0.09$ and $\delta g/g \equiv (g_{22}-g_{11})/g=0.17$, and (c) $g_s/g=0.28$. The color plots display the relative density $\cos\theta$ as a function of position and time, while the insets show the particle current in the majority component across the spin domain during the same time interval. Close to the Manakov limit, 
    the analytical expression \eqref{trajectory} accurately describes the domain's trajectory,
    while as $g_s$ or $\delta g$ increase, oscillations persist in the current and in the position of the domain, and their period is well matched by our prediction. At the same time, the amplitude of the periodic trajectory decreases, so that the configuration comes to mimic a static Josephson junction in the high-$g_s$ limit. Panel (b) shows that the phenomenon is visible and the amplitude and period reasonably close to predictable values with experimentally achievable interaction strengths.}
    \label{fig:evolution_m}
\end{figure*}
This is a periodic function of the momentum, which suggests that if we apply a constant external force, such that the momentum increases linearly in time, the soliton should respond by oscillating. We stress that an adiabatic approximation is involved in this reasoning, which assumes that the application of the external force is able to explore the dispersion relation, i.e., that the Gross-Pitaevskii evolution of an initial soliton state leads to another state within the soliton family. This is valid provided the external potentials vary slowly enough to be approximately constant over the width of the soliton.

The reasoning just outlined raises the question of what the notion of "external force" means in our binary BEC. It is straightforward to show from \eqref{LLE} that the canonical momentum satisfies $\dot{P}=\int \partial_x\omega_D(1-\cos\theta)\dd{x}$. The differential potential $V_1-V_2$ couples to the relative density in the system's dynamical equations, while the sum $V_1+V_2$ couples to the total density, which we have excluded as a dynamical variable. We are thus led, by analogy with Newton's second law, to consider the dynamics of a magnetic soliton under the application of linear potentials such that $\omega_D = \omega_D^0 + \eta x$, with some small gradient $\eta$ \cite{Remark1}.
In this scenario, $N$ remains exactly constant while $\dot{P}=\eta N$: the 
differential potential gradient assumes the role of a constant force, and the momentum increases linearly in time. Within the adiabatic approximation 
we can use Eq. \eqref{imm_energy} to find the evolution of the soliton position $X$ through the relation $\dot{X}=\partial E_{\rm sol}/\partial P$
\cite{Pitaevskii_2016}. This yields
\begin{equation}
\label{trajectory}
X(t) = X(0) + 4\, \frac{ \cos({P(0)}/{2}) - \cos({P(t)}/{2})}{\eta N \sinh(N/2)}.
\end{equation}
This motion corresponds to an adiabatically conserved energy $E=E_{\rm sol}-\eta N X$.
In dimensional units,
the constant force applied to the soliton is $f=N_2
{\rm d}(V_1-V_2)/{\rm d}x = \eta\, N_2 g_s n_0/\xi_s$ and
the period ${\cal T}=2\pi\hbar n_0/f$ is independent of $g_s$. The dimensional amplitude is ${\cal A}=4g_sn_0^3\xi_s/f\sinh(N_2/n_0\xi_s)$, a decreasing function of $g_s$. 

We performed simulations to check our prediction, solving Eq. \eqref{GP} numerically starting from a stationary soliton state obtained from imaginary-time evolution via the procedure detailed in \cite{sartori2013}, and under a potential consisting of a hard wall confining both components to a region much larger than the soliton, supplemented by a linear potential acting on the minority component. For concreteness, and because it is a promising experimental platform, we take the mass to be that of $^{23}$Na. The results of our simulations are illustrated in Fig. \ref{fig:evolution_m}: panels (a) and (c), respectively, illustrate the good quantitative agreement of our predictions for the dynamics close to the Manakov limit and the persistence of the phenomenon and reasonable agreement with predictions for larger $g_s/g$ and $\delta g/g$, which we will discuss shortly. Panel (b) displays the results of a simulation performed with experimentally accessible parameters, using the scattering lengths between the $\ket{F=1,m_F=-1}$ and $\ket{F=2,m_F=-2}$ hyperfine states of $^{23}$Na, demonstrating good agreement with our predictions even in the case where the condition $g_{11}\neq g_{22}$ breaks the mixture's $\mathbb{Z}_2$ symmetry, and presenting evidence that it is possible to observe soliton oscillations in the laboratory.

Previous works have described soliton oscillations like those we predict as Bloch oscillations \cite{Kosevich2001} or attributed them to the periodically changing sign of the soliton's effective mass \cite{Zhao2020}. Although these are appropriate descriptions of a quasiparticle with a periodic dispersion relation, they do not 
explain why the dispersion is periodic in the first place. To do so, it is fruitful to step back from the quasiparticle picture and to consider the soliton as a configuration of a phase-coherent field.
Indeed, the momentum and dispersion relation of this and other solitons are properly defined only by accounting for a global quantity, namely the counterflow momentum (cf. \cite{Jones_1982,Shevchenko1988,Pitaevskii_2016} and \cite{Pitaevskii2016}, Chap. 5).
Once this is done, one finds that the momentum $P$ is proportional to the majority-component phase difference at infinity $\Delta\phi_1 \equiv \phi_1(+\infty)-\phi_1(-\infty)$, since total current conservation enables to express the momentum as $P= \int (\partial_x\varphi - \partial_x\Phi) \dd{x} = 2\Delta\phi_1$. [cf. Eq. \eqref{madelung}]. We now propose what we consider to be a more insightful explanation of magnetic soliton oscillations by explicitly deriving Josephson equations which hold in the slightly immiscible mixture with a magnetic soliton subject to a small uniform differential potential gradient. In this picture, the order parameter subject to Josephson physics is $\psi_1$, with the localized minority component acting, thanks to interspecies repulsion, as a mobile barrier, thus forming a weakly linked junction. The Josephson equation describing the phase across the junction is found by restating $\dot{P}=\eta N$ in terms of the majority-component phase jump (note that $\phi_1$ is approximately constant outside the soliton, so $\Delta \phi_1$ is approximately equal to the phase difference across the soliton):
\begin{equation}
    \label{josephson2}
    \dv{t}\Delta\phi_1 = \frac{1}{2}\eta N.
\end{equation}
The particle current of the majority component across the soliton is $I(t) \equiv \frac{d}{dt} \; \int_{X(t)}^{+\infty} n_1(x,t) \dd{x}= -n_0 
\dot{X}(t)$. We readily obtain from \eqref{trajectory} and \eqref{josephson2}
\begin{equation}
    \label{josephson1}
    I = -I_0 \sin(\Delta\phi_1) 
\end{equation}
where $I_0 = 2 
/\sinh(N/2)$ (independent of $\eta$). Restoring dimensional units gives $I_0 = 2n_0c_s/\sinh(N_2/n_0\xi_s)$.
Equations \eqref{josephson2} and \eqref{josephson1} can be interpreted as the Josephson equations for a junction across which the voltage (or, in the Bose-Josephson picture, the chemical potential difference) is constant and proportional to $\eta$, which is subject to the AC Josephson effect. Note that the apparent complication of a mobile barrier actually simplifies the equations in our regime: the density on either side of the soliton remains constant while the left and right populations change thanks to the fact that the soliton position changes. This means that the only contribution to the chemical potential difference across the junction is that due to the external potential gradient. The constant density also implies that the full Bose-Josephson physics more usually encountered in BECs is not realized; in particular, there is no self-trapping regime in our case. Instead, we have a bosonic system reproducing the physics of a superconducting Josephson junction.

The following picture thus emerges: In the soliton structure we consider, the left and right part of the majority component are separated by the minority component that acts as the analog of a weak link between two superconductors. In such a configuration, an external potential generates a linear increase over time of the phase difference $\Delta\phi_1$ between the right and left ends of the majority component  \cite{Likharev1979}, as described by Eq. \eqref{josephson2}. The current induced in the majority component by this phase mismatch is of course a periodic function of $\Delta\phi_1$ [see Eq. \eqref{josephson1}] and thus of time. The resulting oscillations trigger, by total particle number conservation, similar and opposed oscillations of the minority component, as observed in Fig. \ref{fig:fig1}. The immiscibility of the two components makes the structure particularly robust and the interpretation of the soliton as a Josephson junction does not depend on the precise values of the mixture's interaction parameters.
Thus, although we used the decoupling from the total density dynamics close to the Manakov limit and the mapping to the Landau-Lifshitz equation to treat the problem analytically, we expect oscillations to occur under a constant differential potential gradient even far away from the Manakov limit, as well as for localized spin domains more generally, rather than for solitons specifically. We have confirmed this by solving Eq. \eqref{GP} numerically for increasing values of $g_s$ (we increase $g_{12}$ while keeping $g_{11}$ and $g_{22}$ constant). Examples of the results are shown in Figs. \ref{fig:evolution_m}(b), \ref{fig:evolution_m}(c). We observe an oscillatory trajectory whose period is inversely proportional to the external potential gradient and does not depend strongly on $g_s$ or $\delta g$. 
The particle current remains sinusoidal, with an amplitude decreasing with increasing $g_{s}$, consistent with the fact that this corresponds to a greater energy barrier for the current to tunnel through. While our prediction for the critical current and oscillation amplitude deviate from simulations at higher $g_s$ and $\delta g$, the period continues to match, consistently with the fact that the Josephson frequency does not depend on the characteristics of the junction.

These observations strengthen the proposed Josephson-junction interpretation and reframe the sinusoidal soliton trajectory found in an easy-plane ferromagnet and in a slightly immiscible condensate mixture as a special case of a more general phenomenon. 
In the LLE language, magnetic soliton oscillations in easy-axis ferromagnets are, according to our picture, to be interpreted as manifestations of the spin Josephson effect, with the soliton itself acting as a junction. 
This interpretation is unanticipated from a spintronics
perspective, where nondissipative transport is rather expected for an easy-plane ferromagnet \cite{chen_macdonald_2017,Sonin2010}, but it arises naturally if the phenomenon is realized in an immiscible two-component BEC. In both the ferromagnet and the binary condensate, phase coherence in the order parameter plays the key role. 

We expect an oscillating current to arise in a binary condensate mixture any time a junction is realized, irrespective of the precise parameter values or profile of the initial state. Possible avenues for future research therefore include the effects of the Josephson mechanism under conditions different from the ones considered here, as well as technological applications in atomtronics \cite{roadmap}, such as quantum gyroscopes, where the weak link is implemented by our oscillating minority component.

\nocite{Menotti2002,Gallemi2018}
\begin{acknowledgments}
We thank A. Kamchatnov and G. Lamporesi for fruitful discussions. This work has been funded from Provincia Autonoma di Trento, from the Italian MIUR under the PRIN2017 project CEnTraL (Protocol No. 20172H2SC4). A. Roy acknowledges the support of the Science and Engineering Research Board (SERB), Department of Science and Technology, Government of India under the project SRG/2022/000057 and IIT Mandi seed-grant funds under the project IITM/SG/AR/87. This work has benefited from Q@TN, the joint lab between University of Trento, FBK-Fondazione Bruno Kessler, INFN-National Institute for Nuclear Physics and CNR-National Research Council.\end{acknowledgments}
\bibliography{biblio}

\begin{thebibliography}{57}%
\makeatletter
\providecommand \@ifxundefined [1]{%
 \@ifx{#1\undefined}
}%
\providecommand \@ifnum [1]{%
 \ifnum #1\expandafter \@firstoftwo
 \else \expandafter \@secondoftwo
 \fi
}%
\providecommand \@ifx [1]{%
 \ifx #1\expandafter \@firstoftwo
 \else \expandafter \@secondoftwo
 \fi
}%
\providecommand \natexlab [1]{#1}%
\providecommand \enquote  [1]{``#1''}%
\providecommand \bibnamefont  [1]{#1}%
\providecommand \bibfnamefont [1]{#1}%
\providecommand \citenamefont [1]{#1}%
\providecommand \href@noop [0]{\@secondoftwo}%
\providecommand \href [0]{\begingroup \@sanitize@url \@href}%
\providecommand \@href[1]{\@@startlink{#1}\@@href}%
\providecommand \@@href[1]{\endgroup#1\@@endlink}%
\providecommand \@sanitize@url [0]{\catcode `\\12\catcode `\$12\catcode
  `\&12\catcode `\#12\catcode `\^12\catcode `\_12\catcode `\%12\relax}%
\providecommand \@@startlink[1]{}%
\providecommand \@@endlink[0]{}%
\providecommand \url  [0]{\begingroup\@sanitize@url \@url }%
\providecommand \@url [1]{\endgroup\@href {#1}{\urlprefix }}%
\providecommand \urlprefix  [0]{URL }%
\providecommand \Eprint [0]{\href }%
\providecommand \doibase [0]{https://doi.org/}%
\providecommand \selectlanguage [0]{\@gobble}%
\providecommand \bibinfo  [0]{\@secondoftwo}%
\providecommand \bibfield  [0]{\@secondoftwo}%
\providecommand \translation [1]{[#1]}%
\providecommand \BibitemOpen [0]{}%
\providecommand \bibitemStop [0]{}%
\providecommand \bibitemNoStop [0]{.\EOS\space}%
\providecommand \EOS [0]{\spacefactor3000\relax}%
\providecommand \BibitemShut  [1]{\csname bibitem#1\endcsname}%
\let\auto@bib@innerbib\@empty
%</preamble>
\bibitem [{\citenamefont {Bloch}(1929)}]{bloch1929}%
  \BibitemOpen
  \bibfield  {author} {\bibinfo {author} {\bibfnamefont {F.}~\bibnamefont
  {Bloch}},\ }\href {https://doi.org/10.1007/BF01339455} {\bibfield  {journal}
  {\bibinfo  {journal} {Z. Physik}\ }\textbf {\bibinfo {volume} {52}},\
  \bibinfo {pages} {555} (\bibinfo {year} {1929})}\BibitemShut {NoStop}%
\bibitem [{\citenamefont {Beekman}(2020)}]{beekman2020}%
  \BibitemOpen
  \bibfield  {author} {\bibinfo {author} {\bibfnamefont {A.~J.}\ \bibnamefont
  {Beekman}},\ }\href {https://doi.org/10.1093/ptep/ptaa088} {\bibfield
  {journal} {\bibinfo  {journal} {Prog. Theor. Exp. Phys.}\ }\textbf {\bibinfo
  {volume} {2020}},\ \bibinfo {pages} {073B09} (\bibinfo {year}
  {2020})}\BibitemShut {NoStop}%
\bibitem [{\citenamefont {Avenel}\ and\ \citenamefont
  {Varoquaux}(1988)}]{Avenel1988}%
  \BibitemOpen
  \bibfield  {author} {\bibinfo {author} {\bibfnamefont {O.}~\bibnamefont
  {Avenel}}\ and\ \bibinfo {author} {\bibfnamefont {E.}~\bibnamefont
  {Varoquaux}},\ }\href {https://doi.org/10.1103/PhysRevLett.60.416} {\bibfield
   {journal} {\bibinfo  {journal} {Phys. Rev. Lett.}\ }\textbf {\bibinfo
  {volume} {60}},\ \bibinfo {pages} {416} (\bibinfo {year} {1988})}\BibitemShut
  {NoStop}%
\bibitem [{\citenamefont {Pereverzev}\ \emph {et~al.}(1997)\citenamefont
  {Pereverzev}, \citenamefont {Loshak}, \citenamefont {Backhaus}, \citenamefont
  {Davis},\ and\ \citenamefont {Packard}}]{Pereverzev1997}%
  \BibitemOpen
  \bibfield  {author} {\bibinfo {author} {\bibfnamefont {S.~V.}\ \bibnamefont
  {Pereverzev}}, \bibinfo {author} {\bibfnamefont {A.}~\bibnamefont {Loshak}},
  \bibinfo {author} {\bibfnamefont {S.}~\bibnamefont {Backhaus}}, \bibinfo
  {author} {\bibfnamefont {J.~C.}\ \bibnamefont {Davis}},\ and\ \bibinfo
  {author} {\bibfnamefont {R.~E.}\ \bibnamefont {Packard}},\ }\href
  {https://doi.org/10.1038/41277} {\bibfield  {journal} {\bibinfo  {journal}
  {Nature (London)}\ }\textbf {\bibinfo {volume} {388}},\ \bibinfo {pages}
  {449} (\bibinfo {year} {1997})}\BibitemShut {NoStop}%
\bibitem [{\citenamefont {Sukhatme}\ \emph {et~al.}(2001)\citenamefont
  {Sukhatme}, \citenamefont {Mukharsky}, \citenamefont {Chui},\ and\
  \citenamefont {Pearson}}]{Sukhatme2001}%
  \BibitemOpen
  \bibfield  {author} {\bibinfo {author} {\bibfnamefont {K.}~\bibnamefont
  {Sukhatme}}, \bibinfo {author} {\bibfnamefont {Y.}~\bibnamefont {Mukharsky}},
  \bibinfo {author} {\bibfnamefont {T.}~\bibnamefont {Chui}},\ and\ \bibinfo
  {author} {\bibfnamefont {D.}~\bibnamefont {Pearson}},\ }\href
  {https://doi.org/10.1038/35077024} {\bibfield  {journal} {\bibinfo  {journal}
  {Nature (London)}\ }\textbf {\bibinfo {volume} {411}},\ \bibinfo {pages}
  {280} (\bibinfo {year} {2001})}\BibitemShut {NoStop}%
\bibitem [{\citenamefont {Cataliotti}\ \emph {et~al.}(2001)\citenamefont
  {Cataliotti}, \citenamefont {Burger}, \citenamefont {Fort}, \citenamefont
  {Maddaloni}, \citenamefont {Minardi}, \citenamefont {Trombettoni},
  \citenamefont {Smerzi},\ and\ \citenamefont {Inguscio}}]{Cataliotti2001}%
  \BibitemOpen
  \bibfield  {author} {\bibinfo {author} {\bibfnamefont {F.~S.}\ \bibnamefont
  {Cataliotti}}, \bibinfo {author} {\bibfnamefont {S.}~\bibnamefont {Burger}},
  \bibinfo {author} {\bibfnamefont {C.}~\bibnamefont {Fort}}, \bibinfo {author}
  {\bibfnamefont {P.}~\bibnamefont {Maddaloni}}, \bibinfo {author}
  {\bibfnamefont {F.}~\bibnamefont {Minardi}}, \bibinfo {author} {\bibfnamefont
  {A.}~\bibnamefont {Trombettoni}}, \bibinfo {author} {\bibfnamefont
  {A.}~\bibnamefont {Smerzi}},\ and\ \bibinfo {author} {\bibfnamefont
  {M.}~\bibnamefont {Inguscio}},\ }\href
  {https://doi.org/10.1126/science.1062612} {\bibfield  {journal} {\bibinfo
  {journal} {Science}\ }\textbf {\bibinfo {volume} {293}},\ \bibinfo {pages}
  {843} (\bibinfo {year} {2001})}\BibitemShut {NoStop}%
\bibitem [{\citenamefont {Albiez}\ \emph {et~al.}(2005)\citenamefont {Albiez},
  \citenamefont {Gati}, \citenamefont {F\"olling}, \citenamefont {Hunsmann},
  \citenamefont {Cristiani},\ and\ \citenamefont {Oberthaler}}]{Albiez2005}%
  \BibitemOpen
  \bibfield  {author} {\bibinfo {author} {\bibfnamefont {M.}~\bibnamefont
  {Albiez}}, \bibinfo {author} {\bibfnamefont {R.}~\bibnamefont {Gati}},
  \bibinfo {author} {\bibfnamefont {J.}~\bibnamefont {F\"olling}}, \bibinfo
  {author} {\bibfnamefont {S.}~\bibnamefont {Hunsmann}}, \bibinfo {author}
  {\bibfnamefont {M.}~\bibnamefont {Cristiani}},\ and\ \bibinfo {author}
  {\bibfnamefont {M.~K.}\ \bibnamefont {Oberthaler}},\ }\href
  {https://doi.org/10.1103/PhysRevLett.95.010402} {\bibfield  {journal}
  {\bibinfo  {journal} {Phys. Rev. Lett.}\ }\textbf {\bibinfo {volume} {95}},\
  \bibinfo {pages} {010402} (\bibinfo {year} {2005})}\BibitemShut {NoStop}%
\bibitem [{\citenamefont {Levy}\ \emph {et~al.}(2007)\citenamefont {Levy},
  \citenamefont {Lahoud}, \citenamefont {Shomroni},\ and\ \citenamefont
  {Steinhauer}}]{Levy2007}%
  \BibitemOpen
  \bibfield  {author} {\bibinfo {author} {\bibfnamefont {S.}~\bibnamefont
  {Levy}}, \bibinfo {author} {\bibfnamefont {E.}~\bibnamefont {Lahoud}},
  \bibinfo {author} {\bibfnamefont {I.}~\bibnamefont {Shomroni}},\ and\
  \bibinfo {author} {\bibfnamefont {J.}~\bibnamefont {Steinhauer}},\ }\href
  {https://doi.org/10.1038/nature06186} {\bibfield  {journal} {\bibinfo
  {journal} {Nature (London)}\ }\textbf {\bibinfo {volume} {449}},\ \bibinfo
  {pages} {579} (\bibinfo {year} {2007})}\BibitemShut {NoStop}%
\bibitem [{\citenamefont {Nakata}\ \emph {et~al.}(2014)\citenamefont {Nakata},
  \citenamefont {van Hoogdalem}, \citenamefont {Simon},\ and\ \citenamefont
  {Loss}}]{kouki2014}%
  \BibitemOpen
  \bibfield  {author} {\bibinfo {author} {\bibfnamefont {K.}~\bibnamefont
  {Nakata}}, \bibinfo {author} {\bibfnamefont {K.~A.}\ \bibnamefont {van
  Hoogdalem}}, \bibinfo {author} {\bibfnamefont {P.}~\bibnamefont {Simon}},\
  and\ \bibinfo {author} {\bibfnamefont {D.}~\bibnamefont {Loss}},\ }\href
  {https://doi.org/10.1103/PhysRevB.90.144419} {\bibfield  {journal} {\bibinfo
  {journal} {Phys. Rev. B}\ }\textbf {\bibinfo {volume} {90}},\ \bibinfo
  {pages} {144419} (\bibinfo {year} {2014})}\BibitemShut {NoStop}%
\bibitem [{\citenamefont {Lagoudakis}\ \emph {et~al.}(2010)\citenamefont
  {Lagoudakis}, \citenamefont {Pietka}, \citenamefont {Wouters}, \citenamefont
  {Andr\'e},\ and\ \citenamefont {Deveaud-Pl\'edran}}]{lago2010}%
  \BibitemOpen
  \bibfield  {author} {\bibinfo {author} {\bibfnamefont {K.~G.}\ \bibnamefont
  {Lagoudakis}}, \bibinfo {author} {\bibfnamefont {B.}~\bibnamefont {Pietka}},
  \bibinfo {author} {\bibfnamefont {M.}~\bibnamefont {Wouters}}, \bibinfo
  {author} {\bibfnamefont {R.}~\bibnamefont {Andr\'e}},\ and\ \bibinfo {author}
  {\bibfnamefont {B.}~\bibnamefont {Deveaud-Pl\'edran}},\ }\href
  {https://doi.org/10.1103/PhysRevLett.105.120403} {\bibfield  {journal}
  {\bibinfo  {journal} {Phys. Rev. Lett.}\ }\textbf {\bibinfo {volume} {105}},\
  \bibinfo {pages} {120403} (\bibinfo {year} {2010})}\BibitemShut {NoStop}%
\bibitem [{\citenamefont {Markelov}(1988)}]{Markelov1987}%
  \BibitemOpen
  \bibfield  {author} {\bibinfo {author} {\bibfnamefont {A.~V.}\ \bibnamefont
  {Markelov}},\ }\href
  {http://www.jetp.ras.ru/cgi-bin/e/index/e/67/3/p520?a=list} {\bibfield
  {journal} {\bibinfo  {journal} {Sov. Phys. JETP}\ }\textbf {\bibinfo {volume}
  {67}},\ \bibinfo {pages} {520} (\bibinfo {year} {1988})}\BibitemShut
  {NoStop}%
\bibitem [{\citenamefont {Borovik-Romanov}\ \emph {et~al.}(1988)\citenamefont
  {Borovik-Romanov}, \citenamefont {Bun'kov}, \citenamefont {de~Vaard},
  \citenamefont {Dmitriev}, \citenamefont {Makrotsieva}, \citenamefont
  {Mukharskii},\ and\ \citenamefont {Sergatskov}}]{Borovik1988}%
  \BibitemOpen
  \bibfield  {author} {\bibinfo {author} {\bibfnamefont {A.~S.}\ \bibnamefont
  {Borovik-Romanov}}, \bibinfo {author} {\bibfnamefont {Y.~M.}\ \bibnamefont
  {Bun'kov}}, \bibinfo {author} {\bibfnamefont {A.}~\bibnamefont {de~Vaard}},
  \bibinfo {author} {\bibfnamefont {V.~V.}\ \bibnamefont {Dmitriev}}, \bibinfo
  {author} {\bibfnamefont {V.}~\bibnamefont {Makrotsieva}}, \bibinfo {author}
  {\bibfnamefont {Y.~M.}\ \bibnamefont {Mukharskii}},\ and\ \bibinfo {author}
  {\bibfnamefont {D.~A.}\ \bibnamefont {Sergatskov}},\ }\href
  {http://jetpletters.ru/ps/1095/article_16545.shtml} {\bibfield  {journal}
  {\bibinfo  {journal} {JETP Lett.}\ }\textbf {\bibinfo {volume} {47}},\
  \bibinfo {pages} {478} (\bibinfo {year} {1988})}\BibitemShut {NoStop}%
\bibitem [{\citenamefont {Nogueira}\ and\ \citenamefont
  {Bennemann}(2004)}]{Nogueira2004}%
  \BibitemOpen
  \bibfield  {author} {\bibinfo {author} {\bibfnamefont {F.~S.}\ \bibnamefont
  {Nogueira}}\ and\ \bibinfo {author} {\bibfnamefont {K.-H.}\ \bibnamefont
  {Bennemann}},\ }\href {https://doi.org/10.1209/epl/i2003-10305-x} {\bibfield
  {journal} {\bibinfo  {journal} {{EPL}}\ }\textbf {\bibinfo {volume} {67}},\
  \bibinfo {pages} {620} (\bibinfo {year} {2004})}\BibitemShut {NoStop}%
\bibitem [{\citenamefont {Chen}\ and\ \citenamefont
  {MacDonald}(2017)}]{chen_macdonald_2017}%
  \BibitemOpen
  \bibfield  {author} {\bibinfo {author} {\bibfnamefont {H.}~\bibnamefont
  {Chen}}\ and\ \bibinfo {author} {\bibfnamefont {A.~H.}\ \bibnamefont
  {MacDonald}},\ }\bibinfo {title} {Spin-superfluidity and spin-current
  mediated nonlocal transport},\ in\ \href
  {https://doi.org/10.1017/9781316084366.029} {\emph {\bibinfo {booktitle}
  {Universal Themes of Bose-Einstein Condensation}}},\ \bibinfo {editor}
  {edited by\ \bibinfo {editor} {\bibfnamefont {N.~P.}\ \bibnamefont
  {Proukakis}}, \bibinfo {editor} {\bibfnamefont {D.~W.}\ \bibnamefont
  {Snoke}},\ and\ \bibinfo {editor} {\bibfnamefont {P.~B.}\ \bibnamefont
  {Littlewood}}}\ (\bibinfo  {publisher} {Cambridge University Press},\
  \bibinfo {year} {2017})\ p.\ \bibinfo {pages} {525–548}\BibitemShut
  {NoStop}%
\bibitem [{\citenamefont {Kosevich}\ \emph {et~al.}(1998)\citenamefont
  {Kosevich}, \citenamefont {Gann}, \citenamefont {Zhukov},\ and\ \citenamefont
  {Voronov}}]{Kosevich1998}%
  \BibitemOpen
  \bibfield  {author} {\bibinfo {author} {\bibfnamefont {A.~M.}\ \bibnamefont
  {Kosevich}}, \bibinfo {author} {\bibfnamefont {V.~V.}\ \bibnamefont {Gann}},
  \bibinfo {author} {\bibfnamefont {A.~I.}\ \bibnamefont {Zhukov}},\ and\
  \bibinfo {author} {\bibfnamefont {V.~P.}\ \bibnamefont {Voronov}},\ }\href
  {https://doi.org/10.1134/1.558674} {\bibfield  {journal} {\bibinfo  {journal}
  {J. Exp. Theor. Phys.}\ }\textbf {\bibinfo {volume} {87}},\ \bibinfo {pages}
  {401–407} (\bibinfo {year} {1998})}\BibitemShut {NoStop}%
\bibitem [{\citenamefont {Kosevich}(2001)}]{Kosevich2001}%
  \BibitemOpen
  \bibfield  {author} {\bibinfo {author} {\bibfnamefont {A.~M.}\ \bibnamefont
  {Kosevich}},\ }\href {https://doi.org/10.1063/1.1388415} {\bibfield
  {journal} {\bibinfo  {journal} {Low Temp. Phys.}\ }\textbf {\bibinfo {volume}
  {27}},\ \bibinfo {pages} {513} (\bibinfo {year} {2001})}\BibitemShut
  {NoStop}%
\bibitem [{\citenamefont {Lifshitz}\ and\ \citenamefont
  {Pitaevskii}(1980)}]{Landau_stat2}%
  \BibitemOpen
  \bibfield  {author} {\bibinfo {author} {\bibfnamefont {E.~M.}\ \bibnamefont
  {Lifshitz}}\ and\ \bibinfo {author} {\bibfnamefont {L.~P.}\ \bibnamefont
  {Pitaevskii}},\ }\href@noop {} {\emph {\bibinfo {title} {Statistical Physics,
  Part 2 (Course of Theoretical Physics, volume 9)}}}\ (\bibinfo  {publisher}
  {Pergamon Press},\ \bibinfo {address} {New York},\ \bibinfo {year}
  {1980})\BibitemShut {NoStop}%
\bibitem [{\citenamefont {Zhao}\ \emph {et~al.}(2020)\citenamefont {Zhao},
  \citenamefont {Wang}, \citenamefont {Tang}, \citenamefont {Yang},
  \citenamefont {Yang},\ and\ \citenamefont {Liu}}]{Zhao2020}%
  \BibitemOpen
  \bibfield  {author} {\bibinfo {author} {\bibfnamefont {L.-C.}\ \bibnamefont
  {Zhao}}, \bibinfo {author} {\bibfnamefont {W.}~\bibnamefont {Wang}}, \bibinfo
  {author} {\bibfnamefont {Q.}~\bibnamefont {Tang}}, \bibinfo {author}
  {\bibfnamefont {Z.-Y.}\ \bibnamefont {Yang}}, \bibinfo {author}
  {\bibfnamefont {W.-L.}\ \bibnamefont {Yang}},\ and\ \bibinfo {author}
  {\bibfnamefont {J.}~\bibnamefont {Liu}},\ }\href
  {https://doi.org/10.1103/PhysRevA.101.043621} {\bibfield  {journal} {\bibinfo
   {journal} {Phys. Rev. A}\ }\textbf {\bibinfo {volume} {101}},\ \bibinfo
  {pages} {043621} (\bibinfo {year} {2020})}\BibitemShut {NoStop}%
\bibitem [{\citenamefont {Yu}\ and\ \citenamefont {Blakie}(2022)}]{xiaoquan22}%
  \BibitemOpen
  \bibfield  {author} {\bibinfo {author} {\bibfnamefont {X.}~\bibnamefont
  {Yu}}\ and\ \bibinfo {author} {\bibfnamefont {P.~B.}\ \bibnamefont
  {Blakie}},\ }\href {https://doi.org/10.1103/PhysRevLett.128.125301}
  {\bibfield  {journal} {\bibinfo  {journal} {Phys. Rev. Lett.}\ }\textbf
  {\bibinfo {volume} {128}},\ \bibinfo {pages} {125301} (\bibinfo {year}
  {2022})}\BibitemShut {NoStop}%
\bibitem [{\citenamefont {Schecter}\ \emph {et~al.}(2012)\citenamefont
  {Schecter}, \citenamefont {Gangardt},\ and\ \citenamefont
  {Kamenev}}]{Schecter2012}%
  \BibitemOpen
  \bibfield  {author} {\bibinfo {author} {\bibfnamefont {M.}~\bibnamefont
  {Schecter}}, \bibinfo {author} {\bibfnamefont {D.}~\bibnamefont {Gangardt}},\
  and\ \bibinfo {author} {\bibfnamefont {A.}~\bibnamefont {Kamenev}},\ }\href
  {https://doi.org/https://doi.org/10.1016/j.aop.2011.10.001} {\bibfield
  {journal} {\bibinfo  {journal} {Ann. Phys. (N.Y.)}\ }\textbf {\bibinfo
  {volume} {327}},\ \bibinfo {pages} {639} (\bibinfo {year}
  {2012})}\BibitemShut {NoStop}%
\bibitem [{\citenamefont {Meinert}\ \emph {et~al.}(2017)\citenamefont
  {Meinert}, \citenamefont {Knap}, \citenamefont {Kirilov}, \citenamefont
  {Jag-Lauber}, \citenamefont {Zvonarev}, \citenamefont {Demler},\ and\
  \citenamefont {Nägerl}}]{Meinert2017}%
  \BibitemOpen
  \bibfield  {author} {\bibinfo {author} {\bibfnamefont {F.}~\bibnamefont
  {Meinert}}, \bibinfo {author} {\bibfnamefont {M.}~\bibnamefont {Knap}},
  \bibinfo {author} {\bibfnamefont {E.}~\bibnamefont {Kirilov}}, \bibinfo
  {author} {\bibfnamefont {K.}~\bibnamefont {Jag-Lauber}}, \bibinfo {author}
  {\bibfnamefont {M.~B.}\ \bibnamefont {Zvonarev}}, \bibinfo {author}
  {\bibfnamefont {E.}~\bibnamefont {Demler}},\ and\ \bibinfo {author}
  {\bibfnamefont {H.-C.}\ \bibnamefont {Nägerl}},\ }\href
  {https://doi.org/10.1126/science.aah6616} {\bibfield  {journal} {\bibinfo
  {journal} {Science}\ }\textbf {\bibinfo {volume} {356}},\ \bibinfo {pages}
  {945} (\bibinfo {year} {2017})}\BibitemShut {NoStop}%
\bibitem [{\citenamefont {Schecter}\ \emph {et~al.}(2016)\citenamefont
  {Schecter}, \citenamefont {Gangardt},\ and\ \citenamefont
  {Kamenev}}]{Schecter_2016}%
  \BibitemOpen
  \bibfield  {author} {\bibinfo {author} {\bibfnamefont {M.}~\bibnamefont
  {Schecter}}, \bibinfo {author} {\bibfnamefont {D.~M.}\ \bibnamefont
  {Gangardt}},\ and\ \bibinfo {author} {\bibfnamefont {A.}~\bibnamefont
  {Kamenev}},\ }\href {https://doi.org/10.1088/1367-2630/18/6/065002}
  {\bibfield  {journal} {\bibinfo  {journal} {New J. Phys.}\ }\textbf {\bibinfo
  {volume} {18}},\ \bibinfo {pages} {065002} (\bibinfo {year}
  {2016})}\BibitemShut {NoStop}%
\bibitem [{\citenamefont {Gamayun}\ \emph {et~al.}(2014)\citenamefont
  {Gamayun}, \citenamefont {Lychkovskiy},\ and\ \citenamefont
  {Cheianov}}]{Gamayun2014}%
  \BibitemOpen
  \bibfield  {author} {\bibinfo {author} {\bibfnamefont {O.}~\bibnamefont
  {Gamayun}}, \bibinfo {author} {\bibfnamefont {O.}~\bibnamefont
  {Lychkovskiy}},\ and\ \bibinfo {author} {\bibfnamefont {V.}~\bibnamefont
  {Cheianov}},\ }\href {https://doi.org/10.1103/PhysRevE.90.032132} {\bibfield
  {journal} {\bibinfo  {journal} {Phys. Rev. E}\ }\textbf {\bibinfo {volume}
  {90}},\ \bibinfo {pages} {032132} (\bibinfo {year} {2014})}\BibitemShut
  {NoStop}%
\bibitem [{\citenamefont {Schecter}\ \emph {et~al.}(2015)\citenamefont
  {Schecter}, \citenamefont {Gangardt},\ and\ \citenamefont
  {Kamenev}}]{Schecter2015_com}%
  \BibitemOpen
  \bibfield  {author} {\bibinfo {author} {\bibfnamefont {M.}~\bibnamefont
  {Schecter}}, \bibinfo {author} {\bibfnamefont {D.~M.}\ \bibnamefont
  {Gangardt}},\ and\ \bibinfo {author} {\bibfnamefont {A.}~\bibnamefont
  {Kamenev}},\ }\href {https://doi.org/10.1103/PhysRevE.92.016101} {\bibfield
  {journal} {\bibinfo  {journal} {Phys. Rev. E}\ }\textbf {\bibinfo {volume}
  {92}},\ \bibinfo {pages} {016101} (\bibinfo {year} {2015})}\BibitemShut
  {NoStop}%
\bibitem [{\citenamefont {Gamayun}\ \emph {et~al.}(2015)\citenamefont
  {Gamayun}, \citenamefont {Lychkovskiy},\ and\ \citenamefont
  {Cheianov}}]{Gamayun2015_com}%
  \BibitemOpen
  \bibfield  {author} {\bibinfo {author} {\bibfnamefont {O.}~\bibnamefont
  {Gamayun}}, \bibinfo {author} {\bibfnamefont {O.}~\bibnamefont
  {Lychkovskiy}},\ and\ \bibinfo {author} {\bibfnamefont {V.}~\bibnamefont
  {Cheianov}},\ }\href {https://doi.org/10.1103/PhysRevE.92.016102} {\bibfield
  {journal} {\bibinfo  {journal} {Phys. Rev. E}\ }\textbf {\bibinfo {volume}
  {92}},\ \bibinfo {pages} {016102} (\bibinfo {year} {2015})}\BibitemShut
  {NoStop}%
\bibitem [{\citenamefont {Zibold}\ \emph {et~al.}(2010)\citenamefont {Zibold},
  \citenamefont {Nicklas}, \citenamefont {Gross},\ and\ \citenamefont
  {Oberthaler}}]{Zibold2010}%
  \BibitemOpen
  \bibfield  {author} {\bibinfo {author} {\bibfnamefont {T.}~\bibnamefont
  {Zibold}}, \bibinfo {author} {\bibfnamefont {E.}~\bibnamefont {Nicklas}},
  \bibinfo {author} {\bibfnamefont {C.}~\bibnamefont {Gross}},\ and\ \bibinfo
  {author} {\bibfnamefont {M.~K.}\ \bibnamefont {Oberthaler}},\ }\href
  {https://doi.org/10.1103/PhysRevLett.105.204101} {\bibfield  {journal}
  {\bibinfo  {journal} {Phys. Rev. Lett.}\ }\textbf {\bibinfo {volume} {105}},\
  \bibinfo {pages} {204101} (\bibinfo {year} {2010})}\BibitemShut {NoStop}%
\bibitem [{\citenamefont {Smerzi}\ \emph {et~al.}(1997)\citenamefont {Smerzi},
  \citenamefont {Fantoni}, \citenamefont {Giovanazzi},\ and\ \citenamefont
  {Shenoy}}]{Smerzi1997}%
  \BibitemOpen
  \bibfield  {author} {\bibinfo {author} {\bibfnamefont {A.}~\bibnamefont
  {Smerzi}}, \bibinfo {author} {\bibfnamefont {S.}~\bibnamefont {Fantoni}},
  \bibinfo {author} {\bibfnamefont {S.}~\bibnamefont {Giovanazzi}},\ and\
  \bibinfo {author} {\bibfnamefont {S.~R.}\ \bibnamefont {Shenoy}},\ }\href
  {https://doi.org/10.1103/PhysRevLett.79.4950} {\bibfield  {journal} {\bibinfo
   {journal} {Phys. Rev. Lett.}\ }\textbf {\bibinfo {volume} {79}},\ \bibinfo
  {pages} {4950} (\bibinfo {year} {1997})}\BibitemShut {NoStop}%
\bibitem [{\citenamefont {Raghavan}\ \emph {et~al.}(1999)\citenamefont
  {Raghavan}, \citenamefont {Smerzi}, \citenamefont {Fantoni},\ and\
  \citenamefont {Shenoy}}]{Raghavan1999}%
  \BibitemOpen
  \bibfield  {author} {\bibinfo {author} {\bibfnamefont {S.}~\bibnamefont
  {Raghavan}}, \bibinfo {author} {\bibfnamefont {A.}~\bibnamefont {Smerzi}},
  \bibinfo {author} {\bibfnamefont {S.}~\bibnamefont {Fantoni}},\ and\ \bibinfo
  {author} {\bibfnamefont {S.~R.}\ \bibnamefont {Shenoy}},\ }\href
  {https://doi.org/10.1103/PhysRevA.59.620} {\bibfield  {journal} {\bibinfo
  {journal} {Phys. Rev. A}\ }\textbf {\bibinfo {volume} {59}},\ \bibinfo
  {pages} {620} (\bibinfo {year} {1999})}\BibitemShut {NoStop}%
\bibitem [{\citenamefont {Giovanazzi}\ \emph {et~al.}(2000)\citenamefont
  {Giovanazzi}, \citenamefont {Smerzi},\ and\ \citenamefont
  {Fantoni}}]{Giovanazzi2000}%
  \BibitemOpen
  \bibfield  {author} {\bibinfo {author} {\bibfnamefont {S.}~\bibnamefont
  {Giovanazzi}}, \bibinfo {author} {\bibfnamefont {A.}~\bibnamefont {Smerzi}},\
  and\ \bibinfo {author} {\bibfnamefont {S.}~\bibnamefont {Fantoni}},\ }\href
  {https://doi.org/10.1103/PhysRevLett.84.4521} {\bibfield  {journal} {\bibinfo
   {journal} {Phys. Rev. Lett.}\ }\textbf {\bibinfo {volume} {84}},\ \bibinfo
  {pages} {4521} (\bibinfo {year} {2000})}\BibitemShut {NoStop}%
\bibitem [{\citenamefont {Pitaevskii}\ and\ \citenamefont
  {Stringari}(2016)}]{Pitaevskii2016}%
  \BibitemOpen
  \bibfield  {author} {\bibinfo {author} {\bibfnamefont {L.~P.}\ \bibnamefont
  {Pitaevskii}}\ and\ \bibinfo {author} {\bibfnamefont {S.}~\bibnamefont
  {Stringari}},\ }\href@noop {} {\emph {\bibinfo {title} {Bose-{{Einstein}}
  Condensation and Superfluidity}}},\ International Series of Monographs on
  Physics\ (\bibinfo  {publisher} {{Oxford University Press}},\ \bibinfo
  {address} {{Oxford, United Kingdom}},\ \bibinfo {year} {2016})\BibitemShut
  {NoStop}%
\bibitem [{\citenamefont {Nikuni}\ and\ \citenamefont
  {Williams}(2003)}]{Nikuni2003}%
  \BibitemOpen
  \bibfield  {author} {\bibinfo {author} {\bibfnamefont {T.}~\bibnamefont
  {Nikuni}}\ and\ \bibinfo {author} {\bibfnamefont {J.~E.}\ \bibnamefont
  {Williams}},\ }\href {https://doi.org/10.1023/A:1026206724886} {\bibfield
  {journal} {\bibinfo  {journal} {J. Low Temp. Phys.}\ }\textbf {\bibinfo
  {volume} {133}},\ \bibinfo {pages} {323} (\bibinfo {year}
  {2003})}\BibitemShut {NoStop}%
\bibitem [{\citenamefont {Kim}\ \emph {et~al.}(2017)\citenamefont {Kim},
  \citenamefont {Seo},\ and\ \citenamefont {Shin}}]{Kim2017}%
  \BibitemOpen
  \bibfield  {author} {\bibinfo {author} {\bibfnamefont {J.~H.}\ \bibnamefont
  {Kim}}, \bibinfo {author} {\bibfnamefont {S.~W.}\ \bibnamefont {Seo}},\ and\
  \bibinfo {author} {\bibfnamefont {Y.}~\bibnamefont {Shin}},\ }\href
  {https://doi.org/10.1103/PhysRevLett.119.185302} {\bibfield  {journal}
  {\bibinfo  {journal} {Phys. Rev. Lett.}\ }\textbf {\bibinfo {volume} {119}},\
  \bibinfo {pages} {185302} (\bibinfo {year} {2017})}\BibitemShut {NoStop}%
\bibitem [{\citenamefont {Fava}\ \emph {et~al.}(2018)\citenamefont {Fava},
  \citenamefont {Bienaim\'e}, \citenamefont {Mordini}, \citenamefont {Colzi},
  \citenamefont {Qu}, \citenamefont {Stringari}, \citenamefont {Lamporesi},\
  and\ \citenamefont {Ferrari}}]{Fava2018}%
  \BibitemOpen
  \bibfield  {author} {\bibinfo {author} {\bibfnamefont {E.}~\bibnamefont
  {Fava}}, \bibinfo {author} {\bibfnamefont {T.}~\bibnamefont {Bienaim\'e}},
  \bibinfo {author} {\bibfnamefont {C.}~\bibnamefont {Mordini}}, \bibinfo
  {author} {\bibfnamefont {G.}~\bibnamefont {Colzi}}, \bibinfo {author}
  {\bibfnamefont {C.}~\bibnamefont {Qu}}, \bibinfo {author} {\bibfnamefont
  {S.}~\bibnamefont {Stringari}}, \bibinfo {author} {\bibfnamefont
  {G.}~\bibnamefont {Lamporesi}},\ and\ \bibinfo {author} {\bibfnamefont
  {G.}~\bibnamefont {Ferrari}},\ }\href
  {https://doi.org/10.1103/PhysRevLett.120.170401} {\bibfield  {journal}
  {\bibinfo  {journal} {Phys. Rev. Lett.}\ }\textbf {\bibinfo {volume} {120}},\
  \bibinfo {pages} {170401} (\bibinfo {year} {2018})}\BibitemShut {NoStop}%
\bibitem [{\citenamefont {Lepoutre}\ \emph {et~al.}(2018)\citenamefont
  {Lepoutre}, \citenamefont {Gabardos}, \citenamefont {Kechadi}, \citenamefont
  {Pedri}, \citenamefont {Gorceix}, \citenamefont {Mar\'echal}, \citenamefont
  {Vernac},\ and\ \citenamefont {Laburthe-Tolra}}]{Lepoutre2018}%
  \BibitemOpen
  \bibfield  {author} {\bibinfo {author} {\bibfnamefont {S.}~\bibnamefont
  {Lepoutre}}, \bibinfo {author} {\bibfnamefont {L.}~\bibnamefont {Gabardos}},
  \bibinfo {author} {\bibfnamefont {K.}~\bibnamefont {Kechadi}}, \bibinfo
  {author} {\bibfnamefont {P.}~\bibnamefont {Pedri}}, \bibinfo {author}
  {\bibfnamefont {O.}~\bibnamefont {Gorceix}}, \bibinfo {author} {\bibfnamefont
  {E.}~\bibnamefont {Mar\'echal}}, \bibinfo {author} {\bibfnamefont
  {L.}~\bibnamefont {Vernac}},\ and\ \bibinfo {author} {\bibfnamefont
  {B.}~\bibnamefont {Laburthe-Tolra}},\ }\href
  {https://doi.org/10.1103/PhysRevLett.121.013201} {\bibfield  {journal}
  {\bibinfo  {journal} {Phys. Rev. Lett.}\ }\textbf {\bibinfo {volume} {121}},\
  \bibinfo {pages} {013201} (\bibinfo {year} {2018})}\BibitemShut {NoStop}%
\bibitem [{\citenamefont {Farolfi}\ \emph {et~al.}(2020)\citenamefont
  {Farolfi}, \citenamefont {Trypogeorgos}, \citenamefont {Mordini},
  \citenamefont {Lamporesi},\ and\ \citenamefont {Ferrari}}]{Farolfi2020}%
  \BibitemOpen
  \bibfield  {author} {\bibinfo {author} {\bibfnamefont {A.}~\bibnamefont
  {Farolfi}}, \bibinfo {author} {\bibfnamefont {D.}~\bibnamefont
  {Trypogeorgos}}, \bibinfo {author} {\bibfnamefont {C.}~\bibnamefont
  {Mordini}}, \bibinfo {author} {\bibfnamefont {G.}~\bibnamefont {Lamporesi}},\
  and\ \bibinfo {author} {\bibfnamefont {G.}~\bibnamefont {Ferrari}},\ }\href
  {https://doi.org/10.1103/PhysRevLett.125.030401} {\bibfield  {journal}
  {\bibinfo  {journal} {Phys. Rev. Lett.}\ }\textbf {\bibinfo {volume} {125}},\
  \bibinfo {pages} {030401} (\bibinfo {year} {2020})}\BibitemShut {NoStop}%
\bibitem [{\citenamefont {Kim}\ \emph {et~al.}(2021)\citenamefont {Kim},
  \citenamefont {Hong}, \citenamefont {Lee},\ and\ \citenamefont
  {Shin}}]{Kim2021}%
  \BibitemOpen
  \bibfield  {author} {\bibinfo {author} {\bibfnamefont {J.~H.}\ \bibnamefont
  {Kim}}, \bibinfo {author} {\bibfnamefont {D.}~\bibnamefont {Hong}}, \bibinfo
  {author} {\bibfnamefont {K.}~\bibnamefont {Lee}},\ and\ \bibinfo {author}
  {\bibfnamefont {Y.}~\bibnamefont {Shin}},\ }\href
  {https://doi.org/10.1103/PhysRevLett.127.095302} {\bibfield  {journal}
  {\bibinfo  {journal} {Phys. Rev. Lett.}\ }\textbf {\bibinfo {volume} {127}},\
  \bibinfo {pages} {095302} (\bibinfo {year} {2021})}\BibitemShut {NoStop}%
\bibitem [{\citenamefont {Cominotti}\ \emph {et~al.}(2022)\citenamefont
  {Cominotti}, \citenamefont {Berti}, \citenamefont {Farolfi}, \citenamefont
  {Zenesini}, \citenamefont {Lamporesi}, \citenamefont {Carusotto},
  \citenamefont {Recati},\ and\ \citenamefont {Ferrari}}]{Cominotti2022}%
  \BibitemOpen
  \bibfield  {author} {\bibinfo {author} {\bibfnamefont {R.}~\bibnamefont
  {Cominotti}}, \bibinfo {author} {\bibfnamefont {A.}~\bibnamefont {Berti}},
  \bibinfo {author} {\bibfnamefont {A.}~\bibnamefont {Farolfi}}, \bibinfo
  {author} {\bibfnamefont {A.}~\bibnamefont {Zenesini}}, \bibinfo {author}
  {\bibfnamefont {G.}~\bibnamefont {Lamporesi}}, \bibinfo {author}
  {\bibfnamefont {I.}~\bibnamefont {Carusotto}}, \bibinfo {author}
  {\bibfnamefont {A.}~\bibnamefont {Recati}},\ and\ \bibinfo {author}
  {\bibfnamefont {G.}~\bibnamefont {Ferrari}},\ }\href
  {https://doi.org/10.1103/PhysRevLett.128.210401} {\bibfield  {journal}
  {\bibinfo  {journal} {Phys. Rev. Lett.}\ }\textbf {\bibinfo {volume} {128}},\
  \bibinfo {pages} {210401} (\bibinfo {year} {2022})}\BibitemShut {NoStop}%
\bibitem [{\citenamefont {Son}\ and\ \citenamefont
  {Stephanov}(2002)}]{Son2002}%
  \BibitemOpen
  \bibfield  {author} {\bibinfo {author} {\bibfnamefont {D.~T.}\ \bibnamefont
  {Son}}\ and\ \bibinfo {author} {\bibfnamefont {M.~A.}\ \bibnamefont
  {Stephanov}},\ }\href {https://doi.org/10.1103/PhysRevA.65.063621} {\bibfield
   {journal} {\bibinfo  {journal} {Phys. Rev. A}\ }\textbf {\bibinfo {volume}
  {65}},\ \bibinfo {pages} {063621} (\bibinfo {year} {2002})}\BibitemShut
  {NoStop}%
\bibitem [{\citenamefont {Qu}\ \emph {et~al.}(2016)\citenamefont {Qu},
  \citenamefont {Pitaevskii},\ and\ \citenamefont {Stringari}}]{Qu2016}%
  \BibitemOpen
  \bibfield  {author} {\bibinfo {author} {\bibfnamefont {C.}~\bibnamefont
  {Qu}}, \bibinfo {author} {\bibfnamefont {L.~P.}\ \bibnamefont {Pitaevskii}},\
  and\ \bibinfo {author} {\bibfnamefont {S.}~\bibnamefont {Stringari}},\ }\href
  {https://doi.org/10.1103/PhysRevLett.116.160402} {\bibfield  {journal}
  {\bibinfo  {journal} {Phys. Rev. Lett.}\ }\textbf {\bibinfo {volume} {116}},\
  \bibinfo {pages} {160402} (\bibinfo {year} {2016})}\BibitemShut {NoStop}%
\bibitem [{\citenamefont {Congy}\ \emph {et~al.}(2016)\citenamefont {Congy},
  \citenamefont {Kamchatnov},\ and\ \citenamefont {Pavloff}}]{Congy2016}%
  \BibitemOpen
  \bibfield  {author} {\bibinfo {author} {\bibfnamefont {T.}~\bibnamefont
  {Congy}}, \bibinfo {author} {\bibfnamefont {A.~M.}\ \bibnamefont
  {Kamchatnov}},\ and\ \bibinfo {author} {\bibfnamefont {N.}~\bibnamefont
  {Pavloff}},\ }\href {https://doi.org/10.21468/SciPostPhys.1.1.006} {\bibfield
   {journal} {\bibinfo  {journal} {SciPost Phys.}\ }\textbf {\bibinfo {volume}
  {1}},\ \bibinfo {pages} {006} (\bibinfo {year} {2016})}\BibitemShut {NoStop}%
\bibitem [{\citenamefont {Ivanov}\ \emph {et~al.}(2017)\citenamefont {Ivanov},
  \citenamefont {Kamchatnov}, \citenamefont {Congy},\ and\ \citenamefont
  {Pavloff}}]{Ivanov2017}%
  \BibitemOpen
  \bibfield  {author} {\bibinfo {author} {\bibfnamefont {S.~K.}\ \bibnamefont
  {Ivanov}}, \bibinfo {author} {\bibfnamefont {A.~M.}\ \bibnamefont
  {Kamchatnov}}, \bibinfo {author} {\bibfnamefont {T.}~\bibnamefont {Congy}},\
  and\ \bibinfo {author} {\bibfnamefont {N.}~\bibnamefont {Pavloff}},\ }\href
  {https://doi.org/10.1103/PhysRevE.96.062202} {\bibfield  {journal} {\bibinfo
  {journal} {Phys. Rev. E}\ }\textbf {\bibinfo {volume} {96}},\ \bibinfo
  {pages} {062202} (\bibinfo {year} {2017})}\BibitemShut {NoStop}%
\bibitem [{\citenamefont {Manakov}(1974)}]{Manakov1974}%
  \BibitemOpen
  \bibfield  {author} {\bibinfo {author} {\bibfnamefont {S.~V.}\ \bibnamefont
  {Manakov}},\ }\href
  {http://www.jetp.ras.ru/cgi-bin/e/index/e/38/2/p248?a=list} {\bibfield
  {journal} {\bibinfo  {journal} {Sov. Phys. JETP}\ }\textbf {\bibinfo {volume}
  {38}},\ \bibinfo {pages} {248} (\bibinfo {year} {1974})}\BibitemShut
  {NoStop}%
\bibitem [{\citenamefont {Farolfi}\ \emph {et~al.}(2021)\citenamefont
  {Farolfi}, \citenamefont {Zenesini}, \citenamefont {Trypogeorgos},
  \citenamefont {Mordini}, \citenamefont {Gallem{\'i}}, \citenamefont {Roy},
  \citenamefont {Recati}, \citenamefont {Lamporesi},\ and\ \citenamefont
  {Ferrari}}]{Farolfi2021}%
  \BibitemOpen
  \bibfield  {author} {\bibinfo {author} {\bibfnamefont {A.}~\bibnamefont
  {Farolfi}}, \bibinfo {author} {\bibfnamefont {A.}~\bibnamefont {Zenesini}},
  \bibinfo {author} {\bibfnamefont {D.}~\bibnamefont {Trypogeorgos}}, \bibinfo
  {author} {\bibfnamefont {C.}~\bibnamefont {Mordini}}, \bibinfo {author}
  {\bibfnamefont {A.}~\bibnamefont {Gallem{\'i}}}, \bibinfo {author}
  {\bibfnamefont {A.}~\bibnamefont {Roy}}, \bibinfo {author} {\bibfnamefont
  {A.}~\bibnamefont {Recati}}, \bibinfo {author} {\bibfnamefont
  {G.}~\bibnamefont {Lamporesi}},\ and\ \bibinfo {author} {\bibfnamefont
  {G.}~\bibnamefont {Ferrari}},\ }\href
  {https://doi.org/10.1038/s41567-021-01369-y} {\bibfield  {journal} {\bibinfo
  {journal} {Nat. Phys.}\ }\textbf {\bibinfo {volume} {17}},\ \bibinfo {pages}
  {1359} (\bibinfo {year} {2021})}\BibitemShut {NoStop}%
\bibitem [{\citenamefont {Kosevich}\ \emph {et~al.}(1990)\citenamefont
  {Kosevich}, \citenamefont {Ivanov},\ and\ \citenamefont
  {Kovalev}}]{Kosevich1990}%
  \BibitemOpen
  \bibfield  {author} {\bibinfo {author} {\bibfnamefont {A.}~\bibnamefont
  {Kosevich}}, \bibinfo {author} {\bibfnamefont {B.}~\bibnamefont {Ivanov}},\
  and\ \bibinfo {author} {\bibfnamefont {A.}~\bibnamefont {Kovalev}},\ }\href
  {https://doi.org/https://doi.org/10.1016/0370-1573(90)90130-T} {\bibfield
  {journal} {\bibinfo  {journal} {Phys. Rep.}\ }\textbf {\bibinfo {volume}
  {194}},\ \bibinfo {pages} {117} (\bibinfo {year} {1990})}\BibitemShut
  {NoStop}%
\bibitem [{\citenamefont {Chai}\ \emph {et~al.}(2022)\citenamefont {Chai},
  \citenamefont {You},\ and\ \citenamefont {Raman}}]{Chai2022}%
  \BibitemOpen
  \bibfield  {author} {\bibinfo {author} {\bibfnamefont {X.}~\bibnamefont
  {Chai}}, \bibinfo {author} {\bibfnamefont {L.}~\bibnamefont {You}},\ and\
  \bibinfo {author} {\bibfnamefont {C.}~\bibnamefont {Raman}},\ }\href
  {https://doi.org/10.1103/PhysRevA.105.013313} {\bibfield  {journal} {\bibinfo
   {journal} {Phys. Rev. A}\ }\textbf {\bibinfo {volume} {105}},\ \bibinfo
  {pages} {013313} (\bibinfo {year} {2022})}\BibitemShut {NoStop}%
\bibitem [{sup()}]{supplemental}%
  \BibitemOpen
  \href@noop {} {}\bibinfo {note} {See Supplemental Material which presents the
  parameters relevant for an experimental realisation of the configuration
  discussed in the text, discusses the miscible case, and includes Refs.
  \cite{Menotti2002,Gallemi2018}}\BibitemShut {NoStop}%
\bibitem [{\citenamefont {Matthews}\ \emph {et~al.}(1999)\citenamefont
  {Matthews}, \citenamefont {Anderson}, \citenamefont {Haljan}, \citenamefont
  {Hall}, \citenamefont {Holland}, \citenamefont {Williams}, \citenamefont
  {Wieman},\ and\ \citenamefont {Cornell}}]{Matthews1999}%
  \BibitemOpen
  \bibfield  {author} {\bibinfo {author} {\bibfnamefont {M.~R.}\ \bibnamefont
  {Matthews}}, \bibinfo {author} {\bibfnamefont {B.~P.}\ \bibnamefont
  {Anderson}}, \bibinfo {author} {\bibfnamefont {P.~C.}\ \bibnamefont
  {Haljan}}, \bibinfo {author} {\bibfnamefont {D.~S.}\ \bibnamefont {Hall}},
  \bibinfo {author} {\bibfnamefont {M.~J.}\ \bibnamefont {Holland}}, \bibinfo
  {author} {\bibfnamefont {J.~E.}\ \bibnamefont {Williams}}, \bibinfo {author}
  {\bibfnamefont {C.~E.}\ \bibnamefont {Wieman}},\ and\ \bibinfo {author}
  {\bibfnamefont {E.~A.}\ \bibnamefont {Cornell}},\ }\href
  {https://doi.org/10.1103/PhysRevLett.83.3358} {\bibfield  {journal} {\bibinfo
   {journal} {Phys. Rev. Lett.}\ }\textbf {\bibinfo {volume} {83}},\ \bibinfo
  {pages} {3358} (\bibinfo {year} {1999})}\BibitemShut {NoStop}%
\bibitem [{Rem()}]{Remark1}%
  \BibitemOpen
  \href@noop {} {}\bibinfo {note} {More specifically, we take $V_1=0$ and
  $V_2/g_s n_0=-\eta x$, which is a convenient choice for simulations because
  applying a potential only to the localized minority component decreases
  boundary effects in a finite system. See \cite{supplemental} for a discussion
  of the case $V_1+V_2=0,$ which is in general safer with respect to the
  condition of constant total density, but leads to the introduction of a
  higher-frequency component in the soliton oscillations due to boundary
  effects.}\BibitemShut {Stop}%
\bibitem [{\citenamefont {Pitaevskii}(2016)}]{Pitaevskii_2016}%
  \BibitemOpen
  \bibfield  {author} {\bibinfo {author} {\bibfnamefont {L.~P.}\ \bibnamefont
  {Pitaevskii}},\ }\href {https://doi.org/10.3367/ufne.2016.08.037891}
  {\bibfield  {journal} {\bibinfo  {journal} {Phys.-Usp.}\ }\textbf {\bibinfo
  {volume} {59}},\ \bibinfo {pages} {1028} (\bibinfo {year}
  {2016})}\BibitemShut {NoStop}%
\bibitem [{\citenamefont {Sartori}\ and\ \citenamefont
  {Recati}(2013)}]{sartori2013}%
  \BibitemOpen
  \bibfield  {author} {\bibinfo {author} {\bibfnamefont {A.}~\bibnamefont
  {Sartori}}\ and\ \bibinfo {author} {\bibfnamefont {A.}~\bibnamefont
  {Recati}},\ }\href {https://doi.org/10.1140/epjd/e2013-40635-x} {\bibfield
  {journal} {\bibinfo  {journal} {Eur. Phys. J. D}\ }\textbf {\bibinfo {volume}
  {67}},\ \bibinfo {pages} {260} (\bibinfo {year} {2013})}\BibitemShut
  {NoStop}%
\bibitem [{\citenamefont {Jones}\ and\ \citenamefont
  {Roberts}(1982)}]{Jones_1982}%
  \BibitemOpen
  \bibfield  {author} {\bibinfo {author} {\bibfnamefont {C.~A.}\ \bibnamefont
  {Jones}}\ and\ \bibinfo {author} {\bibfnamefont {P.~H.}\ \bibnamefont
  {Roberts}},\ }\href {https://doi.org/10.1088/0305-4470/15/8/036} {\bibfield
  {journal} {\bibinfo  {journal} {J. Phys. A: Math. Gen.}\ }\textbf {\bibinfo
  {volume} {15}},\ \bibinfo {pages} {2599} (\bibinfo {year}
  {1982})}\BibitemShut {NoStop}%
\bibitem [{\citenamefont {Shevchenko}(1988)}]{Shevchenko1988}%
  \BibitemOpen
  \bibfield  {author} {\bibinfo {author} {\bibfnamefont {S.~I.}\ \bibnamefont
  {Shevchenko}},\ }\href@noop {} {\bibfield  {journal} {\bibinfo  {journal}
  {Sov. J. Low Temp. Phys.}\ }\textbf {\bibinfo {volume} {14}},\ \bibinfo
  {pages} {553} (\bibinfo {year} {1988})},\ \bibinfo {note} {[Fiz. Nizk. Temp.
  {\bf 14}, 1011 (1988)]}\BibitemShut {NoStop}%
\bibitem [{\citenamefont {Likharev}(1979)}]{Likharev1979}%
  \BibitemOpen
  \bibfield  {author} {\bibinfo {author} {\bibfnamefont {K.~K.}\ \bibnamefont
  {Likharev}},\ }\href {https://doi.org/10.1103/RevModPhys.51.101} {\bibfield
  {journal} {\bibinfo  {journal} {Rev. Mod. Phys.}\ }\textbf {\bibinfo {volume}
  {51}},\ \bibinfo {pages} {101} (\bibinfo {year} {1979})}\BibitemShut
  {NoStop}%
\bibitem [{\citenamefont {Sonin}(2010)}]{Sonin2010}%
  \BibitemOpen
  \bibfield  {author} {\bibinfo {author} {\bibfnamefont {E.~B.}\ \bibnamefont
  {Sonin}},\ }\href {https://doi.org/10.1080/00018731003739943} {\bibfield
  {journal} {\bibinfo  {journal} {Adv. Phys.}\ }\textbf {\bibinfo {volume}
  {59}},\ \bibinfo {pages} {181} (\bibinfo {year} {2010})}\BibitemShut
  {NoStop}%
\bibitem [{\citenamefont {Amico}\ \emph {et~al.}(2021)\citenamefont {Amico},
  \citenamefont {Boshier}, \citenamefont {Birkl}, \citenamefont {Minguzzi},
  \citenamefont {Miniatura}, \citenamefont {Kwek}, \citenamefont {Aghamalyan},
  \citenamefont {Ahufinger}, \citenamefont {Anderson}, \citenamefont {Andrei},
  \citenamefont {Arnold}, \citenamefont {Baker}, \citenamefont {Bell},
  \citenamefont {Bland}, \citenamefont {Brantut}, \citenamefont {Cassettari},
  \citenamefont {Chetcuti}, \citenamefont {Chevy}, \citenamefont {Citro},
  \citenamefont {De~Palo}, \citenamefont {Dumke}, \citenamefont {Edwards},
  \citenamefont {Folman}, \citenamefont {Fortagh}, \citenamefont {Gardiner},
  \citenamefont {Garraway}, \citenamefont {Gauthier}, \citenamefont {Günther},
  \citenamefont {Haug}, \citenamefont {Hufnagel}, \citenamefont {Keil},
  \citenamefont {Ireland}, \citenamefont {Lebrat}, \citenamefont {Li},
  \citenamefont {Longchambon}, \citenamefont {Mompart}, \citenamefont {Morsch},
  \citenamefont {Naldesi}, \citenamefont {Neely}, \citenamefont {Olshanii},
  \citenamefont {Orignac}, \citenamefont {Pandey}, \citenamefont
  {Pérez-Obiol}, \citenamefont {Perrin}, \citenamefont {Piroli}, \citenamefont
  {Polo}, \citenamefont {Pritchard}, \citenamefont {Proukakis}, \citenamefont
  {Rylands}, \citenamefont {Rubinsztein-Dunlop}, \citenamefont {Scazza},
  \citenamefont {Stringari}, \citenamefont {Tosto}, \citenamefont
  {Trombettoni}, \citenamefont {Victorin}, \citenamefont {Klitzing},
  \citenamefont {Wilkowski}, \citenamefont {Xhani},\ and\ \citenamefont
  {Yakimenko}}]{roadmap}%
  \BibitemOpen
  \bibfield  {author} {\bibinfo {author} {\bibfnamefont {L.}~\bibnamefont
  {Amico}}, \bibinfo {author} {\bibfnamefont {M.}~\bibnamefont {Boshier}},
  \bibinfo {author} {\bibfnamefont {G.}~\bibnamefont {Birkl}}, \bibinfo
  {author} {\bibfnamefont {A.}~\bibnamefont {Minguzzi}}, \bibinfo {author}
  {\bibfnamefont {C.}~\bibnamefont {Miniatura}}, \bibinfo {author}
  {\bibfnamefont {L.-C.}\ \bibnamefont {Kwek}}, \bibinfo {author}
  {\bibfnamefont {D.}~\bibnamefont {Aghamalyan}}, \bibinfo {author}
  {\bibfnamefont {V.}~\bibnamefont {Ahufinger}}, \bibinfo {author}
  {\bibfnamefont {D.}~\bibnamefont {Anderson}}, \bibinfo {author}
  {\bibfnamefont {N.}~\bibnamefont {Andrei}}, \bibinfo {author} {\bibfnamefont
  {A.~S.}\ \bibnamefont {Arnold}}, \bibinfo {author} {\bibfnamefont
  {M.}~\bibnamefont {Baker}}, \bibinfo {author} {\bibfnamefont {T.~A.}\
  \bibnamefont {Bell}}, \bibinfo {author} {\bibfnamefont {T.}~\bibnamefont
  {Bland}}, \bibinfo {author} {\bibfnamefont {J.~P.}\ \bibnamefont {Brantut}},
  \bibinfo {author} {\bibfnamefont {D.}~\bibnamefont {Cassettari}}, \bibinfo
  {author} {\bibfnamefont {W.~J.}\ \bibnamefont {Chetcuti}}, \bibinfo {author}
  {\bibfnamefont {F.}~\bibnamefont {Chevy}}, \bibinfo {author} {\bibfnamefont
  {R.}~\bibnamefont {Citro}}, \bibinfo {author} {\bibfnamefont
  {S.}~\bibnamefont {De~Palo}}, \bibinfo {author} {\bibfnamefont
  {R.}~\bibnamefont {Dumke}}, \bibinfo {author} {\bibfnamefont
  {M.}~\bibnamefont {Edwards}}, \bibinfo {author} {\bibfnamefont
  {R.}~\bibnamefont {Folman}}, \bibinfo {author} {\bibfnamefont
  {J.}~\bibnamefont {Fortagh}}, \bibinfo {author} {\bibfnamefont {S.~A.}\
  \bibnamefont {Gardiner}}, \bibinfo {author} {\bibfnamefont {B.~M.}\
  \bibnamefont {Garraway}}, \bibinfo {author} {\bibfnamefont {G.}~\bibnamefont
  {Gauthier}}, \bibinfo {author} {\bibfnamefont {A.}~\bibnamefont {Günther}},
  \bibinfo {author} {\bibfnamefont {T.}~\bibnamefont {Haug}}, \bibinfo {author}
  {\bibfnamefont {C.}~\bibnamefont {Hufnagel}}, \bibinfo {author}
  {\bibfnamefont {M.}~\bibnamefont {Keil}}, \bibinfo {author} {\bibfnamefont
  {P.}~\bibnamefont {Ireland}}, \bibinfo {author} {\bibfnamefont
  {M.}~\bibnamefont {Lebrat}}, \bibinfo {author} {\bibfnamefont
  {W.}~\bibnamefont {Li}}, \bibinfo {author} {\bibfnamefont {L.}~\bibnamefont
  {Longchambon}}, \bibinfo {author} {\bibfnamefont {J.}~\bibnamefont
  {Mompart}}, \bibinfo {author} {\bibfnamefont {O.}~\bibnamefont {Morsch}},
  \bibinfo {author} {\bibfnamefont {P.}~\bibnamefont {Naldesi}}, \bibinfo
  {author} {\bibfnamefont {T.~W.}\ \bibnamefont {Neely}}, \bibinfo {author}
  {\bibfnamefont {M.}~\bibnamefont {Olshanii}}, \bibinfo {author}
  {\bibfnamefont {E.}~\bibnamefont {Orignac}}, \bibinfo {author} {\bibfnamefont
  {S.}~\bibnamefont {Pandey}}, \bibinfo {author} {\bibfnamefont
  {A.}~\bibnamefont {Pérez-Obiol}}, \bibinfo {author} {\bibfnamefont
  {H.}~\bibnamefont {Perrin}}, \bibinfo {author} {\bibfnamefont
  {L.}~\bibnamefont {Piroli}}, \bibinfo {author} {\bibfnamefont
  {J.}~\bibnamefont {Polo}}, \bibinfo {author} {\bibfnamefont {A.~L.}\
  \bibnamefont {Pritchard}}, \bibinfo {author} {\bibfnamefont {N.~P.}\
  \bibnamefont {Proukakis}}, \bibinfo {author} {\bibfnamefont {C.}~\bibnamefont
  {Rylands}}, \bibinfo {author} {\bibfnamefont {H.}~\bibnamefont
  {Rubinsztein-Dunlop}}, \bibinfo {author} {\bibfnamefont {F.}~\bibnamefont
  {Scazza}}, \bibinfo {author} {\bibfnamefont {S.}~\bibnamefont {Stringari}},
  \bibinfo {author} {\bibfnamefont {F.}~\bibnamefont {Tosto}}, \bibinfo
  {author} {\bibfnamefont {A.}~\bibnamefont {Trombettoni}}, \bibinfo {author}
  {\bibfnamefont {N.}~\bibnamefont {Victorin}}, \bibinfo {author}
  {\bibfnamefont {W.~v.}\ \bibnamefont {Klitzing}}, \bibinfo {author}
  {\bibfnamefont {D.}~\bibnamefont {Wilkowski}}, \bibinfo {author}
  {\bibfnamefont {K.}~\bibnamefont {Xhani}},\ and\ \bibinfo {author}
  {\bibfnamefont {A.}~\bibnamefont {Yakimenko}},\ }\href
  {https://doi.org/10.1116/5.0026178} {\bibfield  {journal} {\bibinfo
  {journal} {AVS Quantum Science}\ }\textbf {\bibinfo {volume} {3}},\ \bibinfo
  {pages} {039201} (\bibinfo {year} {2021})}\BibitemShut {NoStop}%
\bibitem [{\citenamefont {Menotti}\ and\ \citenamefont
  {Stringari}(2002)}]{Menotti2002}%
  \BibitemOpen
  \bibfield  {author} {\bibinfo {author} {\bibfnamefont {C.}~\bibnamefont
  {Menotti}}\ and\ \bibinfo {author} {\bibfnamefont {S.}~\bibnamefont
  {Stringari}},\ }\href {https://doi.org/10.1103/PhysRevA.66.043610} {\bibfield
   {journal} {\bibinfo  {journal} {Phys. Rev. A}\ }\textbf {\bibinfo {volume}
  {66}},\ \bibinfo {pages} {043610} (\bibinfo {year} {2002})}\BibitemShut
  {NoStop}%
\bibitem [{\citenamefont {Gallem\'{\i}}\ \emph {et~al.}(2018)\citenamefont
  {Gallem\'{\i}}, \citenamefont {Pitaevskii}, \citenamefont {Stringari},\ and\
  \citenamefont {Recati}}]{Gallemi2018}%
  \BibitemOpen
  \bibfield  {author} {\bibinfo {author} {\bibfnamefont {A.}~\bibnamefont
  {Gallem\'{\i}}}, \bibinfo {author} {\bibfnamefont {L.~P.}\ \bibnamefont
  {Pitaevskii}}, \bibinfo {author} {\bibfnamefont {S.}~\bibnamefont
  {Stringari}},\ and\ \bibinfo {author} {\bibfnamefont {A.}~\bibnamefont
  {Recati}},\ }\href {https://doi.org/10.1103/PhysRevA.97.063615} {\bibfield
  {journal} {\bibinfo  {journal} {Phys. Rev. A}\ }\textbf {\bibinfo {volume}
  {97}},\ \bibinfo {pages} {063615} (\bibinfo {year} {2018})}\BibitemShut
  {NoStop}%
\end{thebibliography}%


\begin{thebibliography}{57}%
\makeatletter
\providecommand \@ifxundefined [1]{%
 \@ifx{#1\undefined}
}%
\providecommand \@ifnum [1]{%
 \ifnum #1\expandafter \@firstoftwo
 \else \expandafter \@secondoftwo
 \fi
}%
\providecommand \@ifx [1]{%
 \ifx #1\expandafter \@firstoftwo
 \else \expandafter \@secondoftwo
 \fi
}%
\providecommand \natexlab [1]{#1}%
\providecommand \enquote  [1]{``#1''}%
\providecommand \bibnamefont  [1]{#1}%
\providecommand \bibfnamefont [1]{#1}%
\providecommand \citenamefont [1]{#1}%
\providecommand \href@noop [0]{\@secondoftwo}%
\providecommand \href [0]{\begingroup \@sanitize@url \@href}%
\providecommand \@href[1]{\@@startlink{#1}\@@href}%
\providecommand \@@href[1]{\endgroup#1\@@endlink}%
\providecommand \@sanitize@url [0]{\catcode `\\12\catcode `\$12\catcode
  `\&12\catcode `\#12\catcode `\^12\catcode `\_12\catcode `\%12\relax}%
\providecommand \@@startlink[1]{}%
\providecommand \@@endlink[0]{}%
\providecommand \url  [0]{\begingroup\@sanitize@url \@url }%
\providecommand \@url [1]{\endgroup\@href {#1}{\urlprefix }}%
\providecommand \urlprefix  [0]{URL }%
\providecommand \Eprint [0]{\href }%
\providecommand \doibase [0]{https://doi.org/}%
\providecommand \selectlanguage [0]{\@gobble}%
\providecommand \bibinfo  [0]{\@secondoftwo}%
\providecommand \bibfield  [0]{\@secondoftwo}%
\providecommand \translation [1]{[#1]}%
\providecommand \BibitemOpen [0]{}%
\providecommand \bibitemStop [0]{}%
\providecommand \bibitemNoStop [0]{.\EOS\space}%
\providecommand \EOS [0]{\spacefactor3000\relax}%
\providecommand \BibitemShut  [1]{\csname bibitem#1\endcsname}%
\let\auto@bib@innerbib\@empty
%</preamble>

\bibitem [{\citenamefont {Menotti}\ and\ \citenamefont
  {Stringari}(2002)}]{Menotti2002}%
  \BibitemOpen
  \bibfield  {author} {\bibinfo {author} {\bibfnamefont {C.}~\bibnamefont
  {Menotti}}\ and\ \bibinfo {author} {\bibfnamefont {S.}~\bibnamefont
  {Stringari}},\ }\href {https://doi.org/10.1103/PhysRevA.66.043610} {\bibfield
   {journal} {\bibinfo  {journal} {Phys. Rev. A}\ }\textbf {\bibinfo {volume}
  {66}},\ \bibinfo {pages} {043610} (\bibinfo {year} {2002})}\BibitemShut
  {NoStop}%
\bibitem [{\citenamefont {Kosevich}\ \emph {et~al.}(1990)\citenamefont
  {Kosevich}, \citenamefont {Ivanov},\ and\ \citenamefont
  {Kovalev}}]{Kosevich1990}%
  \BibitemOpen
  \bibfield  {author} {\bibinfo {author} {\bibfnamefont {A.}~\bibnamefont
  {Kosevich}}, \bibinfo {author} {\bibfnamefont {B.}~\bibnamefont {Ivanov}},\
  and\ \bibinfo {author} {\bibfnamefont {A.}~\bibnamefont {Kovalev}},\ }\href
  {https://doi.org/https://doi.org/10.1016/0370-1573(90)90130-T} {\bibfield
  {journal} {\bibinfo  {journal} {Phys. Rep.}\ }\textbf {\bibinfo {volume}
  {194}},\ \bibinfo {pages} {117} (\bibinfo {year} {1990})}\BibitemShut
  {NoStop}%
\bibitem [{\citenamefont {Qu}\ \emph {et~al.}(2016)\citenamefont {Qu},
  \citenamefont {Pitaevskii},\ and\ \citenamefont {Stringari}}]{Qu2016}%
  \BibitemOpen
  \bibfield  {author} {\bibinfo {author} {\bibfnamefont {C.}~\bibnamefont
  {Qu}}, \bibinfo {author} {\bibfnamefont {L.~P.}\ \bibnamefont {Pitaevskii}},\
  and\ \bibinfo {author} {\bibfnamefont {S.}~\bibnamefont {Stringari}},\ }\href
  {https://doi.org/10.1103/PhysRevLett.116.160402} {\bibfield  {journal}
  {\bibinfo  {journal} {Phys. Rev. Lett.}\ }\textbf {\bibinfo {volume} {116}},\
  \bibinfo {pages} {160402} (\bibinfo {year} {2016})}\BibitemShut {NoStop}%
\bibitem [{\citenamefont {Congy}\ \emph {et~al.}(2016)\citenamefont {Congy},
  \citenamefont {Kamchatnov},\ and\ \citenamefont {Pavloff}}]{Congy2016}%
  \BibitemOpen
  \bibfield  {author} {\bibinfo {author} {\bibfnamefont {T.}~\bibnamefont
  {Congy}}, \bibinfo {author} {\bibfnamefont {A.~M.}\ \bibnamefont
  {Kamchatnov}},\ and\ \bibinfo {author} {\bibfnamefont {N.}~\bibnamefont
  {Pavloff}},\ }\href {https://doi.org/10.21468/SciPostPhys.1.1.006} {\bibfield
   {journal} {\bibinfo  {journal} {SciPost Phys.}\ }\textbf {\bibinfo {volume}
  {1}},\ \bibinfo {pages} {006} (\bibinfo {year} {2016})}\BibitemShut {NoStop}%
\bibitem [{\citenamefont {Ivanov}\ \emph {et~al.}(2017)\citenamefont {Ivanov},
  \citenamefont {Kamchatnov}, \citenamefont {Congy},\ and\ \citenamefont
  {Pavloff}}]{Ivanov2017}%
  \BibitemOpen
  \bibfield  {author} {\bibinfo {author} {\bibfnamefont {S.~K.}\ \bibnamefont
  {Ivanov}}, \bibinfo {author} {\bibfnamefont {A.~M.}\ \bibnamefont
  {Kamchatnov}}, \bibinfo {author} {\bibfnamefont {T.}~\bibnamefont {Congy}},\
  and\ \bibinfo {author} {\bibfnamefont {N.}~\bibnamefont {Pavloff}},\ }\href
  {https://doi.org/10.1103/PhysRevE.96.062202} {\bibfield  {journal} {\bibinfo
  {journal} {Phys. Rev. E}\ }\textbf {\bibinfo {volume} {96}},\ \bibinfo
  {pages} {062202} (\bibinfo {year} {2017})}\BibitemShut {NoStop}%
\bibitem [{\citenamefont {Gallem\'{\i}}\ \emph {et~al.}(2018)\citenamefont
  {Gallem\'{\i}}, \citenamefont {Pitaevskii}, \citenamefont {Stringari},\ and\
  \citenamefont {Recati}}]{Gallemi2018}%
  \BibitemOpen
  \bibfield  {author} {\bibinfo {author} {\bibfnamefont {A.}~\bibnamefont
  {Gallem\'{\i}}}, \bibinfo {author} {\bibfnamefont {L.~P.}\ \bibnamefont
  {Pitaevskii}}, \bibinfo {author} {\bibfnamefont {S.}~\bibnamefont
  {Stringari}},\ and\ \bibinfo {author} {\bibfnamefont {A.}~\bibnamefont
  {Recati}},\ }\href {https://doi.org/10.1103/PhysRevA.97.063615} {\bibfield
  {journal} {\bibinfo  {journal} {Phys. Rev. A}\ }\textbf {\bibinfo {volume}
  {97}},\ \bibinfo {pages} {063615} (\bibinfo {year} {2018})}\BibitemShut
  {NoStop}%
\bibitem [{\citenamefont {Pitaevskii}(2016)}]{Pitaevskii_2016}%
  \BibitemOpen
  \bibfield  {author} {\bibinfo {author} {\bibfnamefont {L.~P.}\ \bibnamefont
  {Pitaevskii}},\ }\href {https://doi.org/10.3367/ufne.2016.08.037891}
  {\bibfield  {journal} {\bibinfo  {journal} {Phys.-Usp.}\ }\textbf {\bibinfo
  {volume} {59}},\ \bibinfo {pages} {1028} (\bibinfo {year}
  {2016})}\BibitemShut {NoStop}%
\end{thebibliography}%

\end{document}

\title{Supplemental material to : 
Oscillating solitons and AC Josephson effect in ferromagnetic Bose-Bose mixtures}

\author{S. Bresolin}
\affiliation{INO-CNR BEC Center and Dipartimento di Fisica, Universit\`a di Trento, Via Sommarive 14, 38123 Povo, Trento, Italy}
\author{A. Roy}
\affiliation{INO-CNR BEC Center and Dipartimento di Fisica, Universit\`a di Trento, Via Sommarive 14, 38123 Povo, Trento, Italy}
\affiliation{School of Physical Sciences, Indian Institute of Technology Mandi, Mandi-175075 (H.P.), India}
\author{G. Ferrari}
\affiliation{INO-CNR BEC Center and Dipartimento di Fisica, Universit\`a di Trento, Via Sommarive 14, 38123 Povo, Trento, Italy}
\author{A. Recati}
\email[]{Corresponding Author: alessio.recati@ino.cnr.it}
\affiliation{INO-CNR BEC Center and Dipartimento di Fisica, Universit\`a di Trento, Via Sommarive 14, 38123 Povo, Trento, Italy}
\author{N. Pavloff}
\affiliation{Universit\'e Paris-Saclay, CNRS, LPTMS, 91405, Orsay, France}
\affiliation{Institut Universitaire de France (IUF)}

\maketitle

\section{Values of the parameters used in the numerical simulations}

We present here the parameters used in the simulations whose results are reported in the Letter. The code numerically solves coupled one-dimensional Gross-Pitaevskii equations whose interaction strengths are obtained from three-dimensional values, renormalized as for a cigar-shaped condensate in a harmonic trap with $\omega_y=\omega_z \equiv \omega_\perp \gg \omega_x$.
We denote as $a_\perp=(\hbar/m\omega_\perp)^{1/2}$ the transverse harmonic oscillator length and by $a_{ij}$ the s-wave scattering length characterizing the low-energy 3D interaction between components $i$ and $j$. 
We work in the 1D mean field regime \cite{Menotti2002} where $(a_{ij}/a_\perp)^2\ll n_0 a_{ij}\ll 1$. In this regime $g_{ij}^{1\mathrm{D}} = 2\hbar\omega_\perp a_{ij}$.
The results in Fig. 2(a) of the main text are obtained for $a_{11}=a_{22}=54.5a_0$, $a_{12}=55.1a_0$, where $a_0$ is the Bohr radius. In panel (b) we use $a_{11}=54.5a_0,$ $a_{22}=64.3a_0$(c), while in panel (c) we keep $a_{11}=a_{22}=54.5a_0$ and use $a_{12}=69.6a_0$, and $a_{12}=64.3a_0$, which are the scattering lengths between the hyperfine states $\ket{F=1,m_F=-1}$ and $\ket{F=2,m_F=-2}$ of $^{23}$Na. The total densities at the center of the trap are $n_0^{(a)}=3.3\cdotp 10^9 \rm{m}^{-1}$, $n_0^{(b)}=3.6\cdotp 10^8 \rm{m}^{-1}$, and $n_0^{(c)}=3.3\cdotp 10^8 \rm{m}^{-1}$.

The different potentials acting on the two components can be realized by means of a combined magneto-optical potential whose magnetic and optical parts are both linear. The electric field will exert the same force $f_{\!\scriptscriptstyle E}$ on both components, while the magnetic field will affect the component with larger magnetic moment more strongly. Formally, we can write the external potential contribution to the Hamiltonian as 
\begin{equation*}
 \hat{V}_{\rm{ext}} = f_{\!\scriptscriptstyle B}\,\hat{x} \left ( \ket{1}\bra{1} - 2 \ket{2}\bra{2} \right )+ f_{\!\scriptscriptstyle E}\, \hat{x}(\ket{1}\bra{1} + \ket{2}\bra{2}), 
\end{equation*}
where we use the shorthand notation $\ket{F}\equiv \ket{F,m_F=-F}$ ($F=1$ or 2). 
The condition to obtain $V_1=0,$ $V_2=f \cdot x$ therefore becomes
\begin{equation*}
\begin{cases}
    f_{\!\scriptscriptstyle B} + f_{\!\scriptscriptstyle E} = 0\\
    -2f_{\!\scriptscriptstyle B} + f_{\!\scriptscriptstyle E} = f,
\end{cases}
\end{equation*}
which is solved by $f_{\!\scriptscriptstyle B}=-f_{\!\scriptscriptstyle E}=-f/3$. The magnetic potential gradients used in the simulations correspond, through the expression $V_B = \mu_B |\mathbf{B}|/2$ (where $\mu_B$ is the Bohr magneton) for the magnetic moment potential energy, to magnetic field gradients of 1.1 G/m, 0.9 G/m, and 1.2 G/m respectively.  For such gradients the magnetic field change is very small over the size of the cloud and the scattering lengths are thus practically constant. 

\section{Equal and Opposite Potentials}

\begin{figure*}
    \centering
    \includegraphics[width=\textwidth]{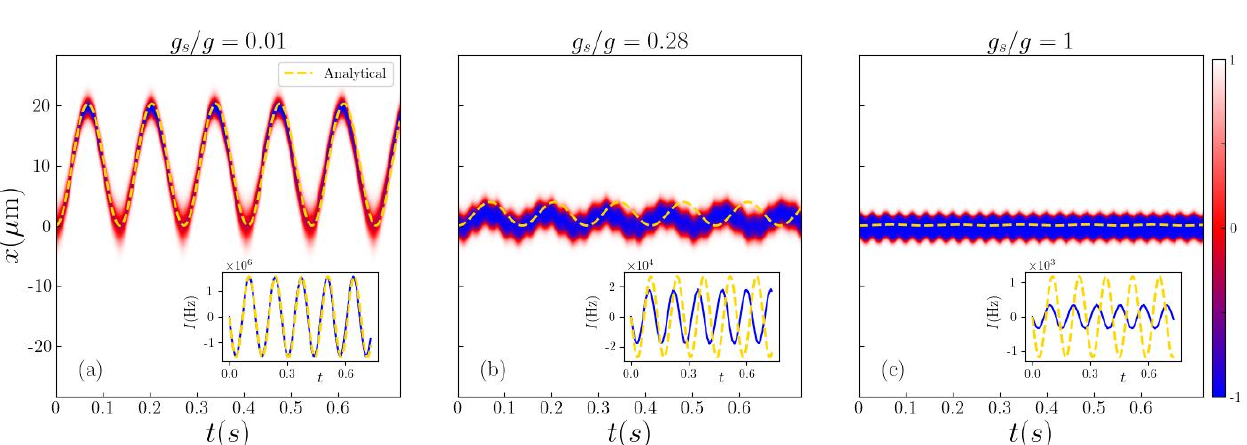}
    \caption{Numerical evolution of a spin domain under a linear differential potential with $V_1=-V_2$ for different values of the interspecies interaction strength. At higher $g_s/g$, the higher-frequency component due to background density oscillations becomes more visible, but the current remains sinusoidal rather than resembling the domain's trajectory.}
    \label{fig:symmetrical}
\end{figure*}

The sum $V_1+V_2$ couples to the total density $n$ in the Lagrangian of the system, and has no equivalent in the Landau-Lifshitz picture, where the magnitude of the magnetization vector is strictly constant. For these reasons, it may seem preferable to use equal and opposite linear potentials to study the dynamics of the condensate mixture, to ensure $V_1+V_2\equiv 0$ and avoid exciting the total density degree of freedom. Actually, as indicated in the main text, a non-zero total potential does not necessarily break the condition of constant density. In fact, using $V_1=-V_2=\eta\, x/2$ complicates the dynamics by acting on the majority component across the entire size of the system, causing the initially flat majority density profile to tilt back and forth periodically, as an effect of the edges of the system. These oscillations affect the motion of the soliton, showing up as a higher-frequency component in its trajectory, as shown in Fig. \ref{fig:symmetrical}. The effect is more noticeable at higher $g_s/g$, since the amplitude of the Josephson oscillations on which the finite-size oscillations are superimposed becomes smaller as the energy barrier grows higher. Thus the trajectory comes to look very different from the sinusoidal curve seen at low $g_s/g$ - but nonetheless the particle current remains perfectly sinusoidal. In this case, a non-sinusoidal soliton trajectory and variations in the background density conspire to keep the behavior of the current in line with the AC Josephson effect, although its period is modified.

\section{Adiabatic soliton motion}
A more detailed view of the approximations we utilize in our analytic treatment can be reached by considering the energy of the system. In a generic configuration of the coupled condensates, with the boundary conditions $n_1\to n_0$ and $n_2\to 0$ at $x\to\pm\infty$, the energy is given by the Gross-Pitaevskii expression
\begin{equation}
\begin{split}
E_{GP} = & \int \Big[ \frac{\hbar^2}{2m} \left(|\partial_x \psi_1|^2 + |\partial_x \psi_2|^2\right) +
V_1(n_1-n_0) \\
& +V_2 n_2 + \frac{g}{2}(n_1^2+n_2^2-n_0^2)
+ g_{12}n_1 n_2 \Big]\dd{x},
\end{split}
\end{equation}
which, in the limit of small and positive $g_{12}-g$ implying constant total density, in adimensional variables and using the $(\theta,\varphi)$ variables of the paper, reduces to 
\begin{equation}
\begin{split}
E = & \int \Big[\tfrac{1}{2} (\partial_x\theta)^2  +
\tfrac{1}{2}\sin^2\theta (1 + (\partial_x \varphi)^2) \\
& + \omega_D(x)(\cos\theta - 1) \Big] \dd{x}.
\end{split}
\end{equation}
This expression corresponds to the energy of an equivalent magnetic system governed by the easy-axis Landau-Lifshitz equation. If $(\theta,\varphi)$ are those of a soliton solution parametrized by the position $X$ and momentum $P$, this yields the energy
\begin{equation}\label{hamilton}
\begin{split}
E & = E_{\rm sol}(P) + \int \omega_D(x)(\cos\theta - 1) \dd{x} \\
 & \approx E_{\rm sol}(P) - \omega_D(X)N,
\end{split}
\end{equation}
where $E_{\rm sol}$ is the soliton's energy [given by Eq. (7) of the paper] and $N = \int (1 - \cos\theta) \dd{x}$ as in the main text. 
The final expression in \eqref{hamilton} gives the Hamiltonian governing the adiabatic motion of the soliton, $X$ and $P$ being the relevant canonical variables. It is obtained in the approximation that $\omega_D(x)$ varies over a typical length scale much larger than the width of the soliton. For linear potentials, this means $\eta \ll 1$. 

It should be noted that, if the external potentials are too large, they will break the condition of constant total density and thus prevent the mapping of the coupled GPEs to a Landau-Lifshitz equation, invalidating our analytical treatment and complicating the interpretation as a Josephson junction. Quantitatively, this means that the difference between the maximum and minimum values of the external potentials should be small compared to the chemical potential. This criterion is 
suggested by analytical treatment of the GPEs and simulations show that if this condition is violated, large density gradients in the initially flat background form and lead to shock waves which quickly break the junction. Numerically, the oscillations are robust up until this happens. Note that, by only applying a potential to the minority component, this condition is weakened, as only the potential difference over the region where the minority density is non-zero is relevant. In this case, the condition becomes equivalent to the one discussed above, imposing $\eta \ll 1$.

\section{The miscible case}
In the main text we focus on the case of immiscible condensate mixtures, corresponding to easy-axis ferromagnets. Analytical expressions for families of solitons also exist for easy-plane ferromagnets \cite{Kosevich1990} and for miscible binary condensates in the small-$|g_s|$ limit where the mapping between Gross-Pitaevskii equation and the Landau Lifshitz equation is applicable \cite{Qu2016,Congy2016,Ivanov2017,Gallemi2018}. These solitons also have a periodic dispersion relation. Thus, similar questions to those we have considered in the Letter can be posed in the miscible (easy-plane) case: can the periodicity in the dispersion relation be attributed to the creation of a mobile Josephson junction? Does this result in an oscillatory current across the soliton when it is subjected to a linear differential potential and is this accompanied by periodic motion of the soliton itself? Does this effect also arise in the generic case of a magnetized domain (as opposed to a true soliton), as it does in immiscible mixtures? Our investigations suggest that some fundamental differences exist between the miscible and immiscible regimes. 

A key difference regards the formation of a mobile Josephson junction. In an immiscible mixture, a region where one component has a high density is able to act as a barrier to the other component. In a miscible mixture, on the other hand, mixing between the two components is favored, so such a configuration can no longer be said to constitute an effective barrier. Rather, the initial barrier is typically unstable and will quickly be destroyed by the mixing of the two condensates (even in the absence of an external force). In the absence of a stable barrier, the argument leading to an interpretation in terms of a Josephson effect is untenable.

\begin{figure}
    \centering
    \includegraphics[width=0.95\linewidth]{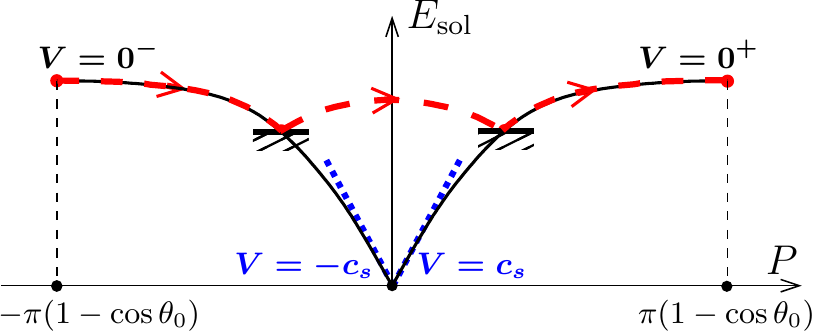}
    \caption{Black solid line: Dispersion relation of the easy-plane soliton for a relative background density $(n_1-n_2)/n_0=\cos\theta_0$. $|P|$ varies between 0 and $\pi(1-\cos\theta_0)$ and $E_{\rm sol}$ between 0 and $2\sin\theta_0-2\theta_0\cos\theta_0$.  The dashed blue lines correspond to the hydrodynamic limit: $E_{\rm sol}=\pm c_s\, P$, with $c_s=\sin\theta_0$. Thick dashed red line: adiabatic motion with bouncing off a hard wall (see the text).}
    \label{fig:misc_dispersion}
\end{figure}

The situation is different when dealing with solitons. In this case, even in the miscible case the object in question is stable and has its own dynamics: a soliton translates unperturbed at constant velocity in a homogeneous system, and in a weakly inhomogeneous system, its motion can be described thanks to a local density approximation 
\cite{Pitaevskii_2016}. 
In the presence of a linear potential acting differently on the two components, this approach runs into some difficulties. One of them is the question of the stability of the soliton under such a force: it may be that, favoring mixing, the force destroys the soliton (see below). Another aspect is the behavior of the easy-plane soliton for increasing velocity: let's assume that a soliton initially at rest is subject to a constant force that drags is toward negative $x$ and linearly increases its momentum\footnote{The soliton is accelerated counter to the force.}. In this case, its representative point $(E_{\rm sol},P)$ in Fig.
\ref{fig:misc_dispersion} will move from the left-most point of the curve ($V=0^-$) down to the origin, following the black dispersion relation. 
When the representative point gets close to the origin, the width of an easy-plane soliton diverges and its amplitude vanishes\footnote{In this instance, the behavior of the easy-plane soliton is similar to that of a dark soliton in a one component condensate, as discussed in a limiting case in Ref. \cite{Gallemi2018}.}. In this situation the adiabatic hypothesis breaks down and the soliton decays into elementary excitations.

A way to avoid this decay is to introduce a hard wall potential which prevents a constant rate of increase of the momentum. In this case, the velocity and momentum of the soliton bouncing off the hard wall are reversed and the soliton changes branch of the dispersion relation, as illustrated schematically by the red dashed line in Fig. \ref{fig:misc_dispersion}. Conservation of energy will let the representative point of the soliton reach the right-most point of the dispersion relation (with zero velocity) and then start over the same downward motion. This is the analog of a ball bouncing off the ground under the effect of the gravitational acceleration.

\begin{figure}[h!]
    \centering
    \includegraphics[width=\linewidth]{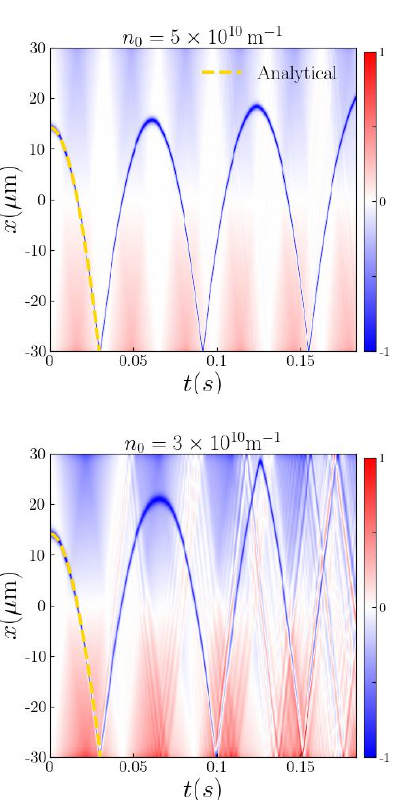}
    \caption{Numerical evolution of a magnetic soliton in a miscible mixture ($g_s/g=-0.01$) under a linear differential potential with gradient $\eta=1.6\times 10^{-3}$ (upper panel) and  $\eta=1.3\times 10^{-2}$ (lower panel). The background relative density is initially zero ($n_1=n_2$: $\cos\theta_0=0$). The soliton in the upper panel is stable on the timescale in which the one in the lower one is destroyed.
    The dashed curves correspond to the analytic prediction \eqref{eq:trajectory}.
    }
    \label{fig:miscible_sol}
\end{figure}
We tested this scenario by running numerical simulations to probe the behavior of solitons in a miscible mixture under a constant external force for various values of several parameters (the relative values of the particle numbers $N_1$ and $N_2$, the total density, the interaction strength $g_s$, the external force, the initial velocity of the soliton) and the results, of which two representative examples are presented in Fig. \ref{fig:miscible_sol}, are compatible with the above scenario: the soliton is initially accelerated counter to the force (towards negative $x$ in our case) until it reaches the wall of the box potential in which the simulations are run. At this point it bounces back and moves in the positive $x$ direction until it stops and starts the same motion again.
We note however, that the adiabatic hypothesis breaks down for forces typically smaller than for easy plane soliton (as already pointed out in \cite{Kosevich1990}). In the presence of a sizeable constant force (i.e., when $\eta$ increases), the velocity of the soliton has become large at the bouncing time and this is associated to a breakdown of adiabaticity: the width of the soliton is large at the point where the gradient of external potential is the largest. In this case, the bouncing is accompanied by a sizeable amount of radiation, as illustrated in the lower panel of Fig. \ref{fig:miscible_sol}. A counter-intuitive phenomenon is then observed: the soliton having lost energy during the bouncing sees its velocity increased and is thus able to reach a point further away from the hard wall than its initial position. This effect  is increased at the next bouncings, eventually leading to a decay of the soliton. This mechanism is clearly at work in the lower panel of Fig. \ref{fig:miscible_sol} and also, although in a less pronounced way, in the upper panel (which corresponds to a lower value of $\eta$).

When the adiabatic approximation holds, one can describe its motion by considering the soliton as a classical particle \cite{Pitaevskii_2016}. The situation is particularly simple in a miscible mixture with equal proportion of the two components, as considered in Fig. \ref{fig:miscible_sol}. In this case the position $X$ of the center of the soliton during its initial motion before bouncing is given by
\begin{equation}
    X(t)=X(0)-\frac{2}{\eta \pi} \Big[1 -\cos(\eta \pi t/2)\Big]
    \; .
    \label{eq:trajectory}
\end{equation}
The good agreement of this prediction with the numerical simulations presented in Fig. \ref{fig:miscible_sol} is a strong support of the above analysis of the ingredients governing the dynamics of an easy-axis soliton.

It is worth emphasising that during our numerical simulations, the miscible mixture proved much more delicate to treat within an adiabatic approximation than the immiscible one. As stated above we attribute this difference to the robustness granted by phase separation to a spin domain in the immiscible phase, thanks to which the dynamics are not very sensitive to the initial condition. In the miscible phase, on the other hand, it is important for the initial state and all subsequent states in a hypothetical adiabatic evolution to truly be solitons. However, we know the exact solitonic solutions only at the demixing transition, whereas for finite $g_s/g,$ these states are only approximate solutions of the Gross-Pitaevskii equation. Decreasing $g_s/g$ makes this discrepancy less significant, but also makes spin excitations softer (as can be seen, for example, from the spin speed of sound $c_s=\sqrt{g_sn_0/2m}$). This means that any external field will excite the spin channel more strongly, making adiabaticity harder to achieve. Correspondingly, it proved necessary in simulations to raise the total density in order to find good agreement with Eq. \eqref{eq:trajectory}.
\bibliography{biblio}